\documentclass[structabstract]{aa}  

\usepackage{color}
\usepackage{natbib}                     % BibTeX style bibliography
\usepackage{graphicx}                   % Pictures and images
\usepackage[varg]{txfonts}              % A&A is printed using the Postscript TX Times-fonts.
\usepackage[english]{babel}             % Language is english
\usepackage{nameref}                    % References
\usepackage{paralist}                   % lists
\usepackage{multirow}

\newcommand{\teff}  {T$_\mathrm{eff}$}
\newcommand{\logg}  {$\log g$}

\begin{document}

  \title{Planets around evolved intermediate-mass stars in open clusters\thanks{
Based on observations collected at the La Silla Observatory, ESO
(Chile), with HARPS/3.6m (runs ID 075.C-0140, 076.C-0429, 078.C-0133, 079.C-0329, 080.C-0071, 081.C-0119, 082.C-0333, 083.C-0413, 091.C-0438, 092.C-0282, 099.C-0304 and 0100.C-0888) and with UVES/VLT 
at the Cerro Paranal Observatory (run 079.C-0131)}}
\subtitle{II. Are there really planets around IC4651No9122, NGC2423No3 and NGC4349No127?}

  \author{E. Delgado Mena\inst{1}
         \and C. Lovis\inst{2}
          \and N. C. Santos\inst{1,3}
          \and J. Gomes da Silva\inst{1}
          \and A. Mortier\inst{4}
          \and M. Tsantaki\inst{5}
          \and S. G. Sousa\inst{1}
          \and P. Figueira\inst{1,6}
          \and M. S. Cunha\inst{1,3}
          \and T. L. Campante\inst{1,3}
          \and V. Adibekyan\inst{1}
          \and J. P. Faria\inst{1}
          \and M. Montalto\inst{7}
        }

  \institute{
    Instituto de Astrof\'isica e Ci\^encias do Espa\c{c}o, Universidade do Porto, CAUP, Rua das Estrelas, 4150-762 Porto, Portugal
    \email{Elisa.Delgado@astro.up.pt}
    \and
    Observatoire de Gen\'eve, Université de Gen\'eve, 51 ch. des Maillettes, 1290 Sauverny, Switzerland
    \and
    Departamento de F\'isica e Astronomia, Faculdade de Ci\^encias, Universidade do Porto, Rua do Campo Alegre, 4169-007 Porto, Portugal
    \and
    Centre for Exoplanet Science, SUPA, School of Physics and Astronomy, University of St Andrews, St Andrews, KY16 9SS, UK
    \and
    Instituto de Radioastronom\'ia y Astrof\'isica, IRyA, UNAM, Campus Morelia, A.P. 3-72, C.P. 58089, Michoac\'an, Mexico
    \and
    European Southern Observatory, Alonso de Cordova 3107, Vitacura, Santiago, Chile
    \and
   Dipartimento di Fisica e Astronomia Galileo Galilei, Universit\'a di Padova, Vicolo dell\'\ Osservatorio 3, I-35122 Padova, Italy
}

  \date{Received date / Accepted date }
% \abstract{}{}{}{}{} 
% 5 {} token are mandatory
 
%----------------------------------------------------------------------------------------
%       Abstract
%----------------------------------------------------------------------------------------
  \abstract
  {}
  {The aim of this work is to search for planets around intermediate-mass stars in open clusters by using data from an extensive survey with more than 15 years of observations.}
  {We obtain high-precision radial velocities (\textit{RV}) with the HARPS spectrograph for a sample of 142 giant stars in 17 open clusters. We fit Keplerian orbits when a significant periodic signal is detected. We also study the variation of stellar activity indicators and line-profile variations to discard stellar-induced signals.}
  {We present the discovery of a periodic \textit{RV} signal compatible with the presence of a planet candidate in the 1.15 Gyr open cluster IC4651 orbiting the 2.06\,M$_\odot$ star  No. 9122. If confirmed, the planet candidate would have a minimum mass of 7.2M\,$_{J}$ and a period of 747 days. However, we also find that the FWHM of the CCF varies with a period close to the \textit{RV}, casting doubts on the planetary nature of the signal. We also provide refined parameters for the previously discovered planet around NGC2423 No. 3 but show evidence that the BIS of the CCF is correlated with the \textit{RV} during some of the observing periods. This fact advises us that this might not be a real planet and that the \textit{RV} variations could be caused by stellar activity and/or pulsations. Finally, we show that the previously reported signal by a brown dwarf around NGC4349 No. 127 is presumably produced by stellar activity modulation.}
  {The long-term monitoring of several red giants in open clusters has allowed us to find periodic \textit{RV} variations in several stars. However, we also show that the follow-up of this kind of stars should last more than one orbital period to detect long-term signals of stellar origin. This work warns that although it is possible to detect planets around red giants, large-amplitude, long-period \textit{RV} modulations do exist in such stars that can mimic the presence of an orbiting planetary body. Therefore, we need to better understand how such \textit{RV} modulations behave as stars evolve along the Red Giant Branch and perform a detailed study of all the possible stellar-induced signals (e.g. spots, pulsations, granulation) to comprehend the origin of \textit{RV} variations.}
  

  \maketitle
%
%_____________________________________________________________________________________________________________________________________________
\section{Introduction}                                  \label{sec:Introduction}

In the last 20 years more than 3500 planets have been discovered, mainly around Main Sequence (MS) solar type stars \cite[exoplanet.eu,][]{schneider11}. One of the most successful methods to detect planets, the radial velocity (\textit{RV}) technique, is more difficult to be used in stars hotter than $\sim$\,6500\,K due to the increase in rotational velocities of those stars and the lack of a sufficient number of spectral lines to determine the velocity shifts in their spectra. Nevertheless, several brown dwarfs and planets have been reported by adapting the \textit{RV} technique to A-F stars \citep{galland06,desort08,lagrange09,borgniet14,borgniet16}. However, in order to understand the planetary formation mechanisms around early F or A stars we need larger samples, together with the determination of planetary masses. Thus, the preferred option to solve this issue has been to apply the \textit{RV} method on K giants, the evolved counterparts of those massive stars, with generally low rotation rates and a larger number of spectral lines in their much cooler spectra when compared with their unevolved MS counterparts \citep[e.g.][]{frink02,sato03,lovis07,niedzielski15}. This allows to probe a different stellar mass range. 

\begin{center}
\begin{table*}
% use packages: array
\caption{Stellar characteristics of the analyzed planet host candidates. Distances and magnitudes are extracted from WEBDA and SIMBAD, respectively.}
\centering
\begin{tabular}{llccc}
\hline
\noalign{\smallskip} 
 & &  IC4651 No. 9122 & NGC2423 No. 3 & NGC4349 No. 127 \\  
\noalign{\smallskip} 
\hline
\hline
\noalign{\smallskip} 
\teff           &  K                & 4720 $\pm$ 71      &   4592 $\pm$ 72        &  4503 $\pm$ 70      \\
\logg           &  (cm\,s$^{-2}$)   & 2.72 $\pm$ 0.15    &   2.33 $\pm$ 0.20      &   1.99 $\pm$ 0.19   \\
$[Fe/H]$        &                   &  0.08 $\pm$ 0.04   &   --0.03 $\pm$ 0.05    &  --0.13 $\pm$ 0.04   \\
$M$             &  M$_\odot$        & 2.06 $\pm$ 0.09    &    2.26 $\pm$ 0.07     &   3.81 $\pm$ 0.23   \\
$R$             &  R$_\odot$        & 8.90 $\pm$ 0.68    &   13.01 $\pm$ 1.11     &  36.98 $\pm$ 4.89  \\
$L$             &  L$_\odot$        & 35.29    &   67.56     &  504.36  \\
Age             & Gyr               & 1.223 $\pm$ 0.150  &   0.850 $\pm$ 0.068    &  0.203 $\pm$ 0.031  \\
Distance        & pc                &  888 &  766  & 2176     \\
$V$             & mag               &  10.91 &  10.04 $\pm$ 0.04  & 10.82 $\pm$ 0.08  \\
$\alpha$        &                   &  17:24:50.1  &  07:37:09.2   & 12:24:35.5   \\
$\delta$        &                   &  --49:56:56.1 &  --13:54:24.0  & --61:49:11.7  \\
\noalign{\medskip} %
\hline
\noalign{\medskip} %
\end{tabular}
\label{tab:stellar}
\end{table*}
\end{center}

A major issue when interpreting \textit{RV} variations in red giants is the presence of intrinsic stellar jitter which shows a typical level of 10-15\,m\,s$^{-1}$ \citep[e.g.][]{setiawan04,hatzes05} and increases towards more evolved stages \citep{hekker08} and for redder stars \citep{frink01}. These short-term (few hours to few days) low-amplitude \textit{RV} variations are mainly caused by radial oscillations, i.e. p-modes \citep[e.g.][]{hatzes07_review}. On the other hand, long-term non-radial pulsations can produce larger \textit{RV} amplitudes of hundreds of m\,s$^{-1}$ \citep[e.g.][]{hatzes99} which eventually can mimic the presence of a planet. Moreover, the combination of several modes of radial oscillations can produce night-to-night variations of $\sim$\,100\,m\,s$^{-1}$ \citep[e.g. see the case of $\alpha$ Boo and $\alpha$ Tau,][]{hatzes93}. The modulation of active regions in red giants can produce large amplitude \textit{RV} variations and on longer timescales as well. Therefore, it is important to carry out long-term observations covering more than one period of the planet candidates (and for a time span larger than the stellar rotational period), to evaluate the stability of the hypothetical planetary signal and its possible relation with the rotational period of the star.

Some interesting correlations between metallicity, stellar mass and the presence of planets have been proposed as more planets have been discovered during the last years. \citet{johnson10} found that the frequency of massive planets is higher around more massive stars but later, \citet{reffert15} reported that the giant planet occurrence rate peaks at $\sim$\,1.9\,M$_\odot$ and then rapidly drops for masses larger than $\sim$\,2.5-3\,M$_\odot$. They also found a clear planet-metallicity correlation for their confirmed planet-host giant star sample. However, the works by \citet{maldonado13} and \citet{mortier13_giants} only found such correlation for giant stars with masses larger than $\sim$\,1.5\,M$_\odot$. 

Therefore, it is clear that in order to understand planet formation mechanisms we need to get accurate masses for the host stars. One of the main problems in characterizing these \textit{evolved} planetary systems is that the determination of masses for red giants is complicated by the crowding of evolutionary tracks with similar stellar parameters in that region of the Hertzsprung-Russell (HR) diagram and thus the determined planetary masses with the \textit{RV} technique may present very large uncertainties \citep{lloyd11,sousa15} \citep[but see ][]{ghezzi15,north17}. As a way to address this issue several programs to discover planets in open clusters have been started. The advantage of working with clusters is that ages and masses of their stars can be better constrained, thus the planetary characterization will be more reliable. Despite the efforts to detect those planets, only few discoveries have been reported. The first planet ever discovered in an open cluster was announced by \citet{sato07}. This was a long period planet (595 days) with a minimum mass of 7.6\,M$_{J}$ orbiting a red giant of 2.7\,M$_\odot$ in the Hyades ($\sim$\,600\,Myr). Soon after, a similar planet was announced around a 2.4\,M$_\odot$ giant in NGC2423 ($m_{2}$ sin \textit{i}\,=\,10.6\,M$_{J}$, P\,=\,750 days) by the \textit{RV} survey around intermediate mass stars of \citet[][hereafter Paper I]{lovis07}. This work also reported the discovery of a brown dwarf orbiting a 3.9\,M$_\odot$ red giant in NGC4349. Finally, \citet{brucalassi14} discovered a Jupiter-like planet in a 120 days orbit around a K giant in M67. We note that the search of planets in open clusters has also targeted MS stars with some detections of Neptune-size planets \citep{meibom13} and hot Jupiters in different clusters  \citep{quinn12,quinn14,malavolta16,brucalassi14,brucalassi16}. 

The aim of this work is to present the new results of our \textit{RV} survey started in Paper I which includes the discovery of a long-period planet candidate in IC4651. The outline of the paper is as follows: in Sect. \ref{sec:observations} we present the data and derivation of stellar parameters. The planet candidate found in IC4651 is debated in Sect. \ref{sec:IC4651} and the discussion of the signals previously atributed to the presence of a planet in NGC2423 is presented in Sect. \ref{sec:NGC2423}. In Sect. \ref{sec:NGC4349} we provide evidence that the previously discovered signal from a brown dwarf in NGC4349 is most likely caused by modulation of stellar magnetic activity. Finally, in Sects. \ref{sec:discussion} and  \ref{sec:conclusion} we present a general discussion and conclusions of the results.

\section{Observations and sample} \label{sec:observations}

The \textit{RV} sample used in this work is fully described in Paper I. The objects analyzed here have been followed for nearly 5 years (from March 2005 to October 2009) with HARPS (ESO-3.6m, La Silla) and some of them were also observed with CORALIE (1.2\,m-Swiss Telescope, La Silla) in previous years. In summary, the survey has focused in open clusters observable from La Silla, with bright giants (maximum \textit{V}\,=\,10 or 13, for each instrument respectively) having masses between 1.5 and 4\,M$_\odot$. Moreover, only clusters with at least 3 giants known to be non-binary cluster members were chosen. Additional observations have been taken during 2017 and 2018 for the targets showing large \textit{RV} variations. The observing programs are detailed in a footnote in page 1. The observations were done using classical fibre spectroscopy mode (no simultaneous calibration) and the exposures times were estimated in order to have individual spectra with a signal-to-noise ratio (S/N) \,$\sim$\,30. This gives a typical \textit{RV} photon-noise of $\sim$\,3.5\,m\,s$^{-1}$ which is enough to detect massive planets around the surveyed stars.

In total, 142 stars were monitored within 17 open clusters using the HARPS spectrograph at the ESO 3.6m telescope (\textit{R}\,$\sim$\,115000). The stellar parameters, namely the effective temperature (\teff), surface gravity (\logg), metallicity ([Fe/H]), and  microturbulence ($\xi_{t}$) for most of the clusters were presented in \citet{santos09,santos12} and improved upon in \citet{delgado16} (by using higher S/N data and a Fe linelist optimized for cool stars) together with the derivation of stellar ages, masses and radii. In this paper we will focus on the results for three stars in the clusters IC4651, NGC2423 and NGC4349. The parameters for those targets are summarized in Table \ref{tab:stellar}.

\section{A planet candidate around IC4651 No. 9122} \label{sec:IC4651}

\subsection{Parent star characteristics} 

%----------------------------------------- FIGURE ---------------------------------------------------
\begin{figure}%[h]
\centering
\includegraphics[width=8cm]{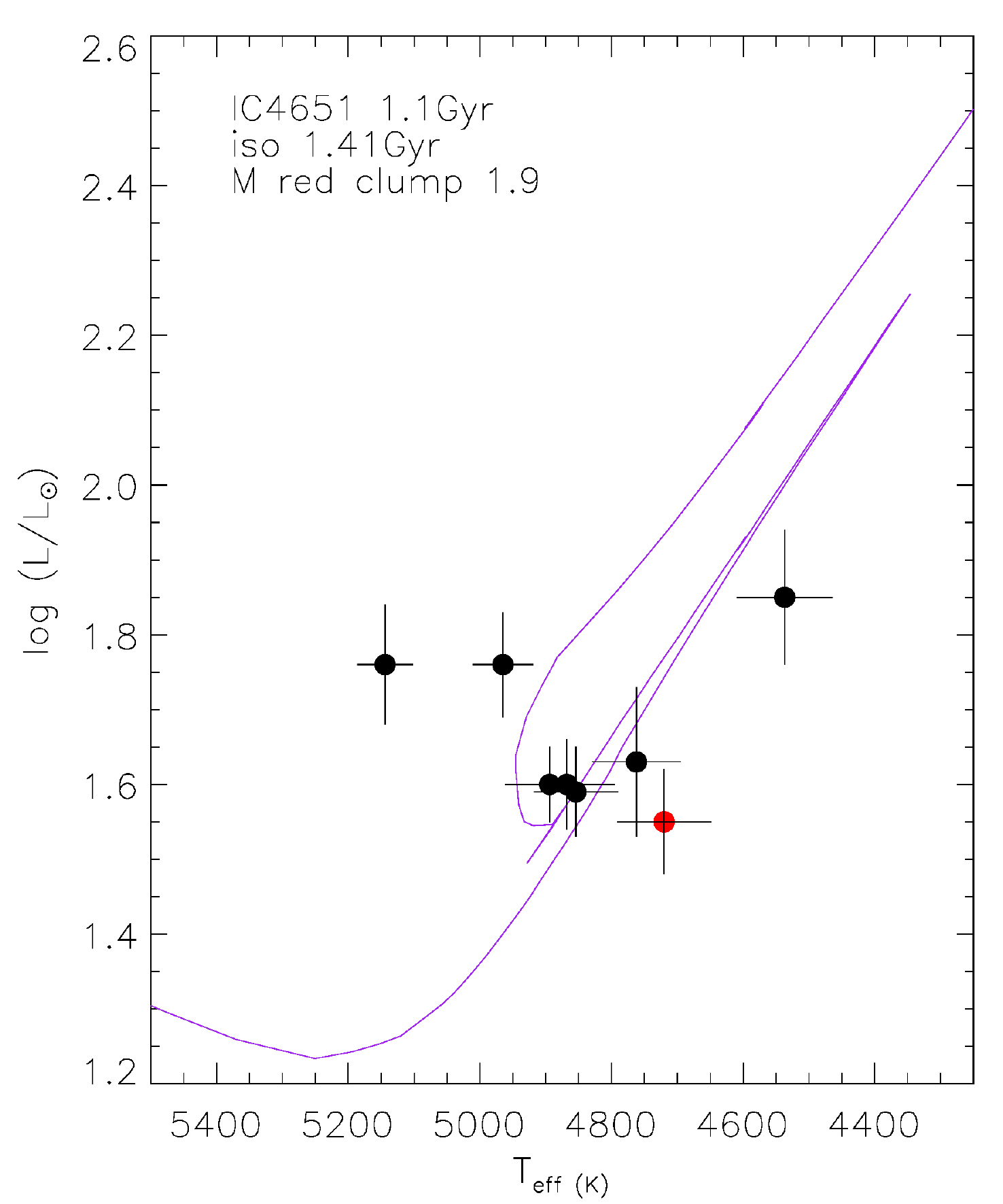}
\caption{HR diagram for the cluster IC4651. The purple line is the isochrone of 1.25 Gyr and Z\,=\,0.018 ([Fe/H]\,=\,0.09 dex). The red dot is the planet host IC4651 No. 9122.} 
\label{IC4651_HR}
\end{figure}
%--------------------------------------------------------------------------------------------------

Our sample contains 8 giant stars in the open cluster IC4651 (distance \textit{d}\,=\,888 pc) with an average metallicity [Fe/H]\,=\,0.06\,$\pm$\,0.06 dex \citep[see Table B.1 of][]{delgado16}. In Fig. \ref{IC4651_HR} we show the HR diagram for this cluster. We use PARSEC isochrones \citep{bressan12} to derive masses and ages \citep[see Table B.2 of][]{delgado16}. The isochrone which better matches our stars has an age of 1.41 Gyr, slightly higher than the 1.14\,Gyr provided by the WEBDA database\footnote{http://www.univie.ac.at/webda/}. Our planet candidate host, with \teff\,=\,4720\,$\pm$\,71\,K, \logg\,=\,2.72\,$\pm$\,0.15\,dex, M\,=\,2.06\,$\pm$\,0.09\,M$_\odot$ and R\,=\,8.90\,$\pm$\,0.68\,R$_\odot$ (see Table \ref{tab:stellar}) seems to be on the first ascent of the Red Giant Branch (RGB) while most of the other stars seem to be located in the red clump. The mean radial velocity of the giants in this cluster is -30.28\,$\pm$\,1.33\,km\,s$^{-1}$ while the mean \textit{RV} of IC4651 No. 9122 is -30.26\,km\,s$^{-1}$, thus this star is likely a cluster member. We estimated a \textit{v} sin \textit{i} of 3.89\,km\,s$^{-1}$ in \citet{delgado16} for a fixed macroturbulence velocity of 2.36\,km\,s$^{-1}$ \citep[with the empirical formula by][]{valenti05}. This leads to a maximum rotational period of $\sim$115 days. If we consider the empirical relations for macrotubulence given by \citet{gray05} and \citet{hekker07} for class III stars we would obtain a macroturbulence velocity of 5.25\,km\,s$^{-1}$ which would then cause \textit{v} sin \textit{i} to decrease to 0.1\,km\,s$^{-1}$ (the lowest limit accepted by the procedure to derive \textit{v} sin \textit{i}). This is a probably unrealistic value which would lead to a rotational period of 4503 days.

In order to illustrate the stellar jitter of the observations we plot the \textit{RV} variations of the stars in the cluster in Fig.\,\ref{IC4651_scatter}. Most of the stars exhibit a scatter around 13\,m\,s$^{-1}$ while our planet candidate host shows a dispersion of 60\,m\,s$^{-1}$, clearly above the average stellar jitter. Thus, this may hint to the presence of a planet. The star with the \textit{RV} dispersion of 26\,m\,s$^{-1}$ is IC4651 No. 9791, which show \textit{RV} peak to peak variations of 80\,m\,s$^{-1}$. The photometric survey by \citet{sahay14} reported that this star is a long-period variable but could not establish its period due to the limited duration of the observations (130 days). Therefore, it is probable that the variability observed in brightness is related with the scatter in \textit{RV} and of stellar origin, although it is not possible either to determine any period with our \textit{RV} data. Interestingly, other two stars in this cluster, No. 8540 and No. 9025 also show brightness variability as reported by \citet{sahay14}, which in the case of the latter reaches 0.4 magnitudes with a 73 days cadence. However, the \textit{RV} scatter for these two stars is lower, 10.8\,m\,s$^{-1}$ and 12.1\,m\,s$^{-1}$, respectively, of the level of the expected noise in these stars. To diagnose these effects we will analyse different activity proxies in Section 3.4.

\subsection{The radial velocity dataset} 

In the upper panel of Fig.\,\ref{ts_9122_additional}, the \textit{RV} data spanning 12.5 years of observations is shown. We collected a total of 47 points between ESO periods 75 and 83. These data already allow to infer the possible presence of a planet but we found hints that the \textit{RV} variability might be of stellar origin. Therefore, we collected 4 \textit{RV} additional points\footnote{We note that we have applied a small negative offset to the \textit{RV} points taken after 2015, when an upgrade of the HARPS fibers took place \citep{locurto15}. The shift was calculated by extrapolating the measurements of Table 3 in \citet{locurto15} to the FWHM of the spectra for this star.} during periods 99 and 100 (between April and October 2017). Moreover we found in the ESO archive 6 extra points taken during 2013 and 2014 (programs 091.C-0438 and 092.C-0282 by Sanzia Alves et al.) as listed in Table \ref{IC4651_rv}. The typical photon noise dominated errors in \textit{RV} of the data is $\sim$\,3\,m\,s$^{-1}$. However, the \textit{RV} variations are usually dominated by the stellar jitter which is much higher in this kind of stars as discussed in previous subsection. For example, solar-like p-mode oscillations in this star are expected to produce an amplitude of $\sim$\,4\,m\,s$^{-1}$ and a period of 0.13 days estimated with the scaling relations of \citet{kjeldsen95}. 
Since the average scatter in the \textit{RV} data (for the 'stable' stars) is close to 15\,m\,s$^{-1}$ we decided to add quadratically this noise to the photon noise before fitting the data to determine the planet orbit. We used the \textit{yorbit} algorithm \cite{segransan11} to fit the whole dataset with a model composed of a Keplerian function. \textit{Yorbit} uses a hybrid method based on a fast linear algorithm (Levenberg-Marquardt) and genetic operators (breeding,  mutations, crossover), and has been optimised to explore the parameter space for Keplerian fitting of radial velocity datasets. In our case, the global search for orbital parameters was made with a genetic algorithm.

\begin{figure}
\centering
\includegraphics[width=1\linewidth]{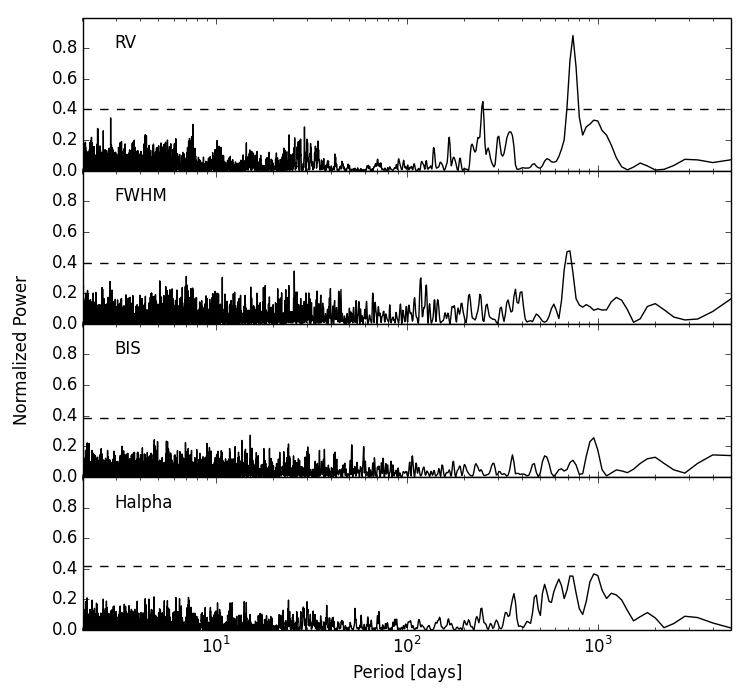}
\caption{Generalized Lomb-Scargle periodograms of \textit{RV}, FWHM, BIS and H$\alpha$ index for IC4651 No. 9122. The dashed line indicates the FAP at 1\% level.} 
\label{per_9122_additional}
\end{figure}

\begin{figure}[h]
\centering
\includegraphics[width=1\linewidth]{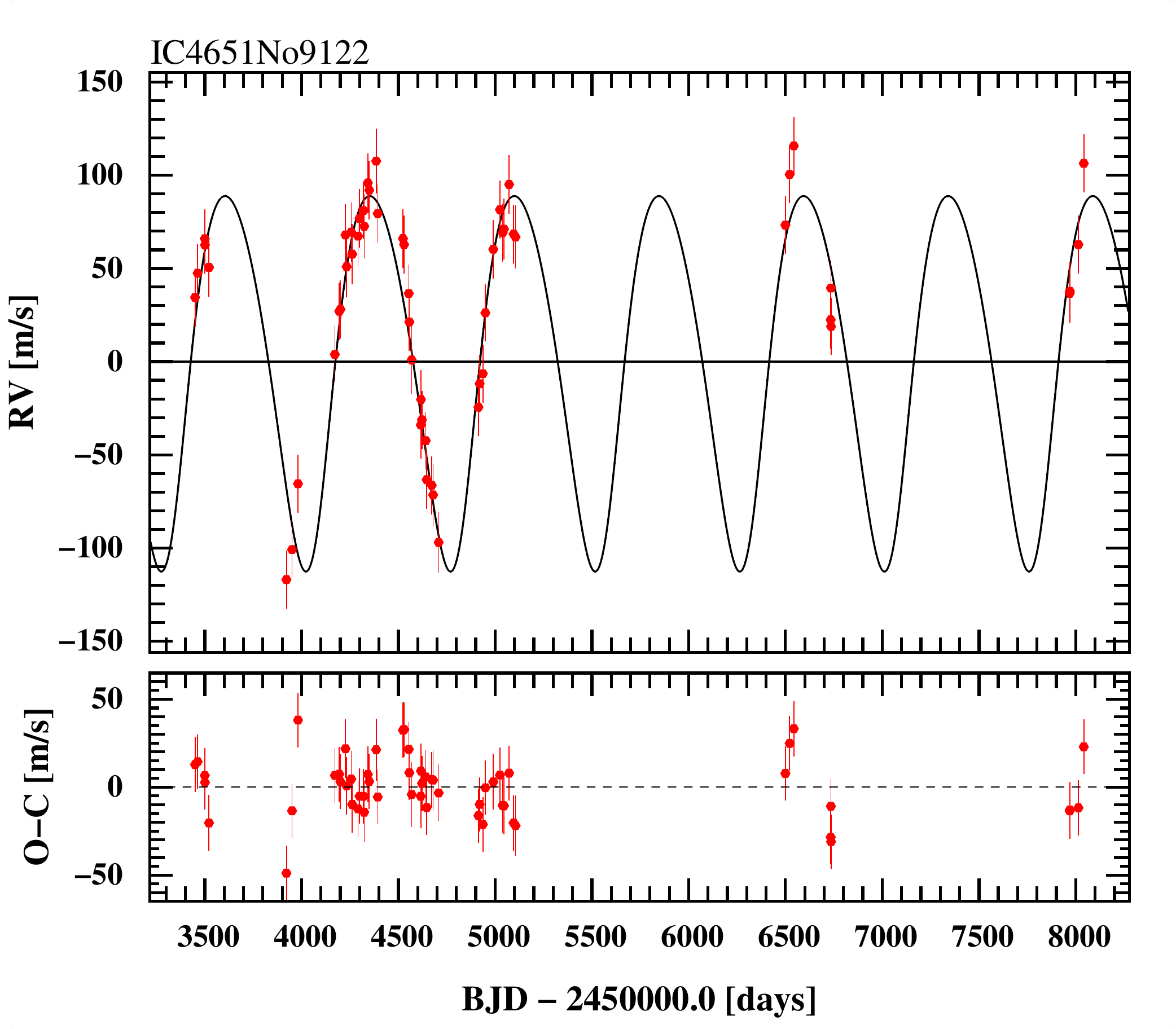}
\caption{Radial velocity curve as a function of time for IC4651 No. 9122. The fitted orbit corresponds to a period of 747 days. A stellar jitter of 15\,m\,s$^{-1}$ has been added to the error bars.} 
\label{IC4651all_phase}
\end{figure}

An analysis using Generalised Lomb-Scargle (GLS) periodograms \citep[e.g.][]{zechmeister09} was performed for \textit{RV} (see upper panel of Fig.\,\ref{per_9122_additional}). The false-alarm-probability (FAP) was computed by bootstrapping the data and statistically significant peaks were considered for values above FAP = 1\%. A significant signal with a period of $\sim$741 days can be clearly observed in the periodogram. This periodic variation is well fitted by a Keplerian function with $P$\,=\,747\,days, $K$\,=\,100.8\,m\,s$^{-1}$, and $e$\,=\,0.15 (see Table \ref{tab:orbital_additional}). These values correspond to the expected signal induced by a planet with 7.2\,M$_{J}$ in a 2.05 AU semi-major orbit. The phase curve of the best fitted solution can be seen in Fig.\,\ref{IC4651all_phase}. The dispersion of the residuals is 17\,m\,s$^{-1}$ and the reduced $\chi^{2}$ is 1.34. As a further test we run a Markov chain Monte Carlo (MCMC) to evaluate the white noise in our data. We get orbital parameters similar to those found by \textit{yorbit} and we determine a jitter of 13.5\,m\,s$^{-1}$ which is very close to the average \textit{RV} in the cluster. Therefore, we can conclude that the assumption of adding a noise of 15\,m\,s$^{-1}$ does not alter the results and it is an acceptable solution to treat the stellar-induced noise that cannot be properly modelled with our current data.

\begin{figure}%[h]
\centering
\includegraphics[width=1\linewidth]{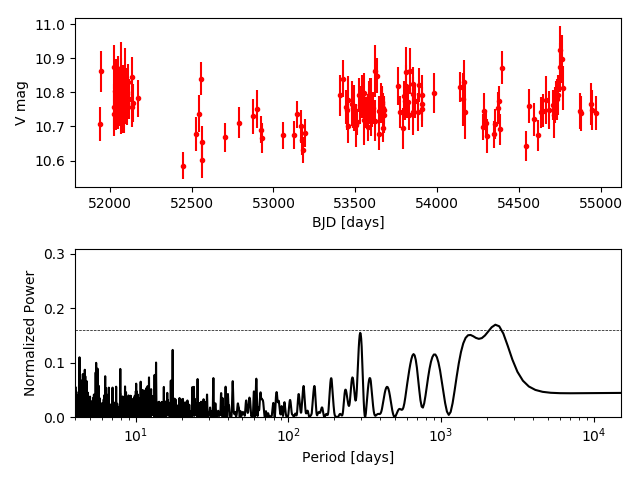}
\caption{GLS of V magnitude for IC4651No9122 using ASAS data. The dashed line indicates the FAP at 1\% level.} 
\label{IC4651_phot}
\end{figure}

\begin{figure}%[h]
\centering
\includegraphics[width=1\linewidth]{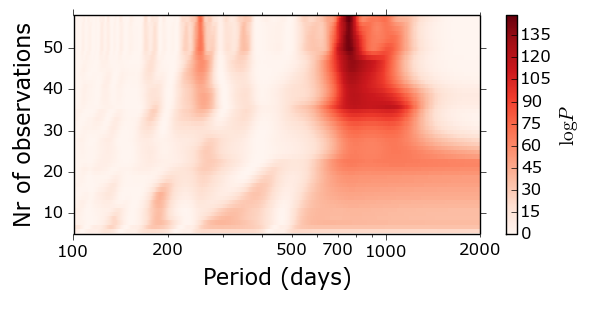}
\caption{Stacked periodogram for \textit{RV} measurements in IC4651 No. 9122.} 
\label{stacked_IC4651}
\end{figure}

\subsection{Photometry} 
This star has been observed within the All Sky Automated Survey (ASAS) from the las Campanas Observatory (Chile) with observations available between 2001 and 2009 \citep{pojmanski02}, hence contemporaneous to our first set of \textit{RV} measurements. We found 151 measurements classified as good quality (they are given the grade A or B, with average errors of 0.05 mag) which show a $\sim$0.3 peak-to-peak variability in V magnitude. We downloaded the light curves in V magnitude from the The ASAS-3 Photometric V-band Catalogue\footnote{http://www.astrouw.edu.pl/asas/?page=aasc$\&$catsrc=asas3} and performed a GLS to detect any possible variability with the same period as the planet candidate. In Fig. \ref{IC4651_phot} we can see that there are two signals just above the FAP = 1\% line with periods of 296 and 2273 days. Since we cannot constrain well the rotational period of the star it is difficult to claim whether any of these signals in the photometry correspondo to such period.

\subsection{Stellar activity and line profile analysis} 
To infer about activity modulations that could interfere with \textit{RV} we measured a simultaneous activity index based on the H$\alpha$ line. We choose not to use the \ion{Ca}{ii} H\&K and \ion{Na}{i} D$_1$ \& D$_2$ lines due to very low S/N ($<$\,3) on the former and line contamination on the latter. The H$\alpha$ index was calculated as in \citet{gomesdasilva11} and implemented using ACTIN\footnote{ACTIN is a FITS extractor and Activity Index calculator tool implemented in Python and available at https://github.com/gomesdasilva/ACTIN.}. The flux in the H$\alpha$ line was measured at 6562.81 \AA\,with a 1.6 \AA\,bandpass. Two reference lines were measured using two 10.75 \AA\,and 8.75 \AA\,windows centred at 6550.87 \AA\,and 6580.31 \AA, respectively. We then divided the flux in the H$\alpha$ line by the two reference lines to obtain the index. The errors on the fluxes were computed using photon noise, $\sqrt{N}$, with $N$ being the total flux in the band, and the error on the index was computed via propagation of errors. To further investigate \textit{RV} variations caused by stellar atmospheric phenomena, we also used the full-width-at-half-maximum (FWHM) of the cross-correlation function (CCF). These values and their errors are provided by the HARPS pipeline. Moreover, we also analysed the Bisector Inverse Slope \citep[BIS,][]{queloz01} of the CCF but we note that this diagnostic of line asymmetry loses sensitivity for very low stellar rotations such as the case we are studying here \citep[e.g.][]{saar98,santos03_planet,queloz09,santos14}.

A periodogram analysis was carried out for FWHM, BIS and H$\alpha$ index to infer about possible impact of stellar activity or long-term oscillations on the observed \textit{RV} (see Fig. \ref{per_9122_additional}). The H$\alpha$ periodogram shows two peaks at 952 and 714\,days but they are not statistically significant. Interestingly, there are also peaks at 952\,days in the BIS and \textit{RV} periodograms although they are not significant. On the other hand, the FWHM periodogram also shows a long-period signal at $\sim$714\,days, above the FAP = 1\% line. This signal is probably due to rotational modulation of active regions in the atmosphere \citep[see e.g.][]{larson93,lambert87} since it matches the same period as H$\alpha$ index. As explained in Sect. 3.1, the determination of the rotation period is a very degenerate problem (depending on the assumption for macroturbulent velocity) and it is difficult then to determine whether that signal in the FWHM periodogram is caused by spot modulation or not. Nevertheless, we note that it has been reported that stars can have long-term activity cycles on top of the typical rotation-modulated cycle \citep{dumusque11,lovis11}. The fact that the period of the FWHM lies so close to the period of the planet candidate (which in turn matches one of the peaks of the H$\alpha$ index periodogram) casts doubts on the planetary hypothesis and will be further discussed in Sect. \ref{sec:discussion}. We note that during the referee process of this paper the work by \citet{leao18} announced the presence of a planet around IC4651 No. 9122 using the same \textit{RV} data presented here. However, in that work the variation of the FWHM of the CCF was not explored which made the authors conclude that the signal should be caused by a planet. Our analysis casts doubts on the planetary nature of the periodic signal and shows the importance of analysing all possible stellar indicators.

Stellar activity and pulsations are also expected to cause deformations in the line profile of spectral lines. Therefore, we also explored any possible correlation between the \textit{RV} and several line-profile indicators such as FWHM, Bisector Inverse Slope \citep[BIS,][]{queloz01}, V$_{span}$ \citep{boisse11}, biGauss \citep{nardetto06} and V$_{asy}$, BIS+ and BIS- \citep{figueira13} by using the python code \textit{LineProf.py}\footnote{https://bitbucket.org/pedrofigueira/line-profile-indicators} developed by \citet{figueira13}. We did not find any strong correlation for any of the indicators, with a maximum value of the Pearson correlation coefficient of 0.3 for the relation of V$_{span}$ with \textit{RV}.

To further evaluate the significance of the periodic signals for \textit{RV} we calculated the Stacked Bayesian general Lomb-Scargle (BGLS) periodogram which is shown in Fig. \ref{stacked_IC4651}. This tool, developed by \citet{mortier17}, aims at tracking the S/N of the detection of a signal over time. If an orbiting body, such as a planet, is the cause of the observed periodic signal, we expect that the signal gets more significant as we add more observations. On the other hand, stellar activity signals are unstable and incoherent and thus the Stacked BLGS will show that the probability of having a signal at a given period varies (decreasing or increasing) when adding more data. In Fig. \ref{stacked_IC4651} we can see that the peak at $\sim$700\,days gets more significant by adding more data and stabilises after $\sim$\,50 observations with some small fluctuations. This would be the expected behaviour for the signature of a planetary companion.

\begin{figure}%[h]
\centering
\includegraphics[width=1\linewidth]{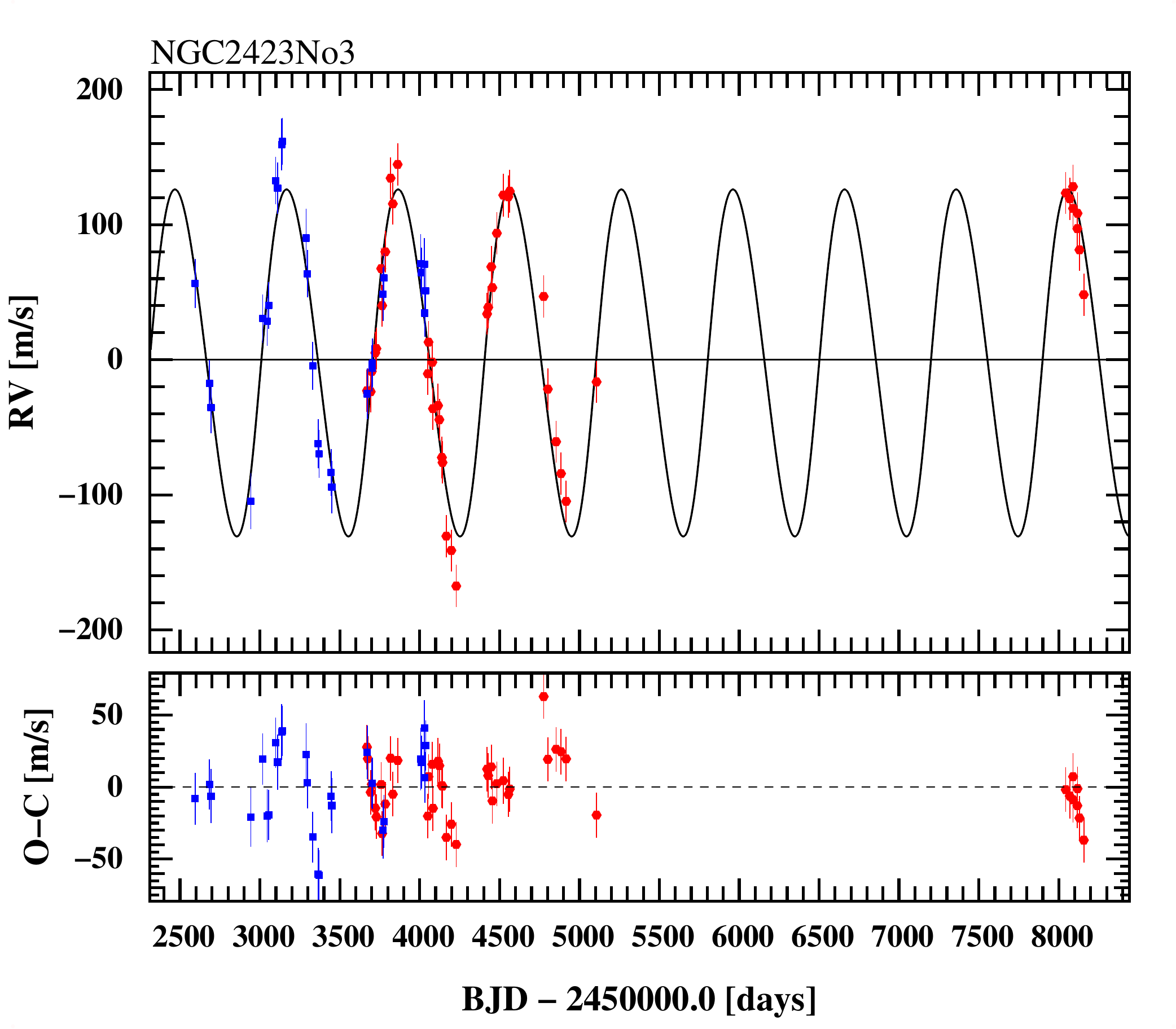}
\caption{Radial velocity curve as a function of time for NGC2423 No. 3 using the complete dataset. The fitted orbit corresponds to a period of 698 days. A stellar jitter of 15\,m\,s$^{-1}$ has been added to the error bars.} 
\label{NGC2423all_phase}
\end{figure}

\section{Is the \textit{RV} variation of NGC2423 No. 3 due to the presence of a planet?} \label{sec:NGC2423}

In Paper I, showing the first results of this survey, a planet in the cluster NGC2423 was announced after collecting 46 \textit{RV} points (28 with CORALIE and 18 with HARPS) during 1529 days. The reported planet had a mass of 10.6\,M$_{J}$ orbiting around NGC2423 No. 3 with a period of 714.3 days in an orbit of eccentricity 0.21 and semimajor axis 2.1\,AU. The parent star has a mass of 2.26$\pm$0.07\,M$_\odot$ and seems to be ascending the RGB and it is probably the most evolved star among the surveyed ones within this cluster. Most of the stars in this cluster exhibit a \textit{RV} scatter below 30\,m\,s$^{-1}$ while NGC2423 No. 3 clearly constrasts with a dispersion of 72\,m\,s$^{-1}$. We refer the reader to Paper I for the HR diagram and \textit{RV} jitter histogram for this cluster. The parameters of NGC2423 No. 3 are detailed in Table \ref{tab:stellar}. For this star we can also estimate the rotational period from the projected rotational velocity. If we consider a macroturbulence velocity of 2.15\,m\,s$^{-1}$ \citep[with the formula by][]{gray05} we obtain \textit{v} sin \textit{i}\,=\,3.84\,km\,s$^{-1}$. Given that the radius of this star is 13\,R$_\odot$ we would obtain a maximum period of 172 days. On the other hand, if we consider the empirical relation of \cite{hekker07}, the macrotubulence velocity would be 5\,km\,s$^{-1}$ which leads to a very small \textit{v} sin \textit{i} (0.1\,km\,s$^{-1}$) which in turn corresponds to a maximum rotational period of 6627 days.

\begin{figure}%[h]
\centering
\includegraphics[width=1\linewidth]{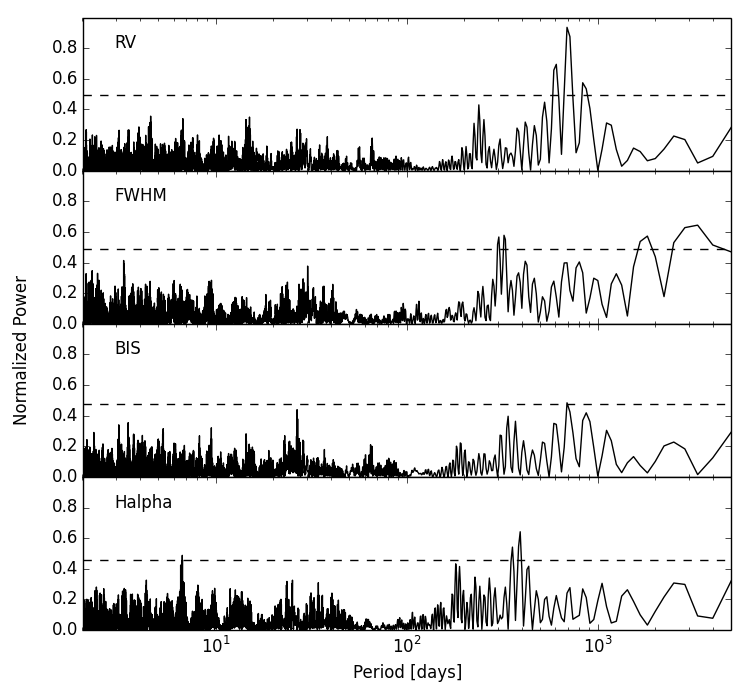}
\caption{Generalized Lomb-Scargle periodograms of \textit{RV}, FWHM, BIS and H$\alpha$ index for NGC2423 No. 3. The dashed line indicates the FAP at 1\% level.} 
\label{per_3_additional}
\end{figure}

\begin{figure}%[h]
\centering
\includegraphics[width=1\linewidth]{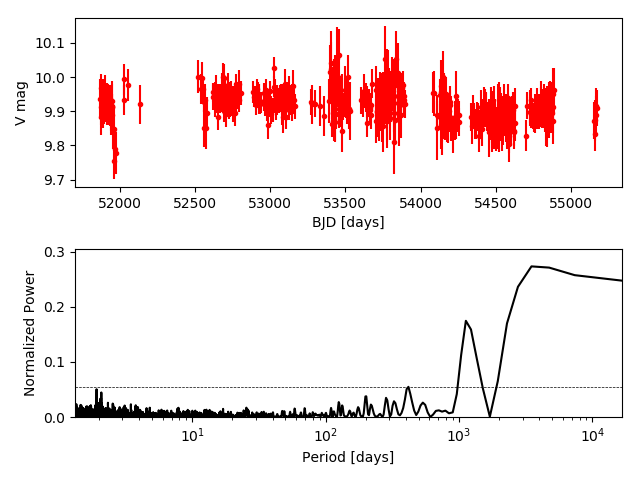}
\caption{GLS of V magnitude for NGC2423No3 using ASAS data. The dashed line indicates the FAP at 1\% level.} 
\label{NGC2423_phot}
\end{figure}

\subsection{The radial velocity dataset}

In Fig. \ref{NGC2423all_phase} we show the \textit{RV} data for NGC2423 No. 3 spanning more than 15 years of observations (73 observations, 45 of them with HARPS) and the orbital solution obtained with \textit{yorbit} for a planet with a period of 698.6 days in a circular orbit with \textit{a}\,=\,2.02 AU. The typical photon noise dominated errors in \textit{RV} of the data is $\sim$\,3\,m\,s$^{-1}$ and $\sim$\,10\,m\,s$^{-1}$ for HARPS and CORALIE data, respectively. We note that we have also added here an error of 15\,m\,s$^{-1}$ before fitting the data\footnote{In Paper I an error of 20\,m\,s$^{-1}$ was added, which was shown later to be overestimated, thus we decided to use here 15\,m\,s$^{-1}$ as done for the previous star.}. By using the updated stellar mass (2.26$\pm$0.07\,M$_\odot$) from \citet{delgado16} we derive a planetary mass $m_{2}$ sin \textit{i}\,=\,9.6\,M$_{J}$ (see Table \ref{tab:orbital_additional}). The period and semi-major axis are very similar to previous values but the eccentricity is significantly lower (the former value was 0.21). The reduced $\chi^{2}$ of the fit is 2.13 and the dispersion of the residuals is 22.4\,m\,s$^{-1}$, a value similar to the typical error bars in \textit{RV} considering the stellar jitter. The residuals shown in Fig. \ref{NGC2423all_phase} seem to have a periodic behaviour. A GLS periodogram of the residuals shows two significant peaks around 400\,days. Interestingly, this is close to the period of 417\,days seen in the photometry (see below). However, we cannot properly fit such residuals with a single keplerian orbit or a sinusoidal function. We note here that by doing a two Keplerian fit, for the highest amplitude signal we would obtain similar orbital parameters as for the single Keplerian while the secondary signal would present a very bad fit. This probably reflects the fact that the residuals we see in the single Keplerian are of stellar origin, caused by different processes with different periods that our data cadence cannot resolve and thus cannot be fitted.

\subsection{Photometry} 
We collected 497 photometric measurements from the ASAS-3 catalogue (with greade A or B) taken between 2000 and 2009 with average errors of 0.04 mag. The lightcurve shows a $\sim$0.15 peak-to-peak variability in V magnitude. After applying a GLS we can see in Fig. \ref{NGC2423_phot} a clear peak above the FAP = 1\% line with a period of 1124 days. Moreover, there is a second peak just above the FAP line at 417 days. Neither of these periods is close to the \textit{RV} period but they could be related to the rotation of the star.

\begin{center}
\begin{table*}
% use packages: array
\caption{Orbital and physical parameters for the planet candidates.}
\centering
\begin{tabular}{llccc}
\hline
\noalign{\smallskip} 
 & &  IC4651 No. 9122 & NGC2423 No. 3  & NGC4349 No. 127  \\  
\noalign{\smallskip} 
\hline
\hline
\noalign{\smallskip} 
$\gamma$           & [km s$^{-1}$]      & --30.2868 $\pm$ 0.0107 &  18.3014 $\pm$ 0.0078      & --11.4322 $\pm$ 0.0119 \\
P                  & [days]             &   747.22 $\pm$ 2.95    &  698.61 $\pm$ 2.72         & 671.94 $\pm$   5.32 \\
K                  & [m s$^{-1}$]       &   100.765 $\pm$ 8.427  &  128.451 $\pm$ 5.851       & 229.454 $\pm$     9.932  \\
\textit{e}         &                    &    0.150 $\pm$ 0.068   &  0.088 $\pm$ 0.041         & 0.046 $\pm$   0.033 \\
$\omega$           & [deg]              &  --141.9 $\pm$ 16.1    &  --102.1 $\pm$ 23.2        & 168.3 $\pm$  54.5  \\
Tp                 & [BJD-2\,400\,000]  & 55575.85 $\pm$ 36.31   &  55083.27 $\pm$ 43.16      & 54306.62 $\pm$ 100.64 \\
$m_{2}$ sin \textit{i} & [M$_{J}$]          &      7.202             &     9.621              & 24.097           \\ 
\textit{a}         & [AU]               &      2.05              &      2.02                  & 2.35             \\ 
\noalign{\medskip} %                                                                     
\hline                                                                                   
\noalign{\medskip} %                                                                            
N$_{meas}$         &                    &      57                &    73                      &  46 \\
Span               & [days]             &      4592              &    5563                    &  1587 \\
$\Delta$v (HARPS-Coralie) & [km s$^{-1}$] &    ---               &    0.007                   &  --- \\ 
N$\sigma$          & [m s$^{-1}$]       &     17.00              &   22.40                    &  31.79 \\
\noalign{\medskip} %                                                                       
\hline
\noalign{\medskip} %
\end{tabular}
\label{tab:orbital_additional}
\end{table*}
\end{center}

\subsection{Stellar activity and line profile analysis}
In order to find any sign of stellar origin in the \textit{RV} variation for this star we performed a similar analysis as for IC4651 No. 9122. In Figs. \ref{ts_3_additional} and \ref{per_3_additional} we show the time series and periodograms of the HARPS data only. The quality and resolution of CORALIE data does not allow to investigate the BIS for this star. For this star we could measure the \ion{Na}{i} D$_1$ \& D$_2$ lines but they do not show any periodic variation. The \textit{RV} periodogram (Fig. \ref{per_3_additional}) shows a strong peak at $\sim$689.8\,days, similar to the period found with the Keplerian fit. The FWHM periodogram presents a significant peak at $\sim$\,322\,days with an alias at $\sim$\,303\,days caused by the long gap in time between the two sets of observations\footnote{We note that we have corrected the FWHM values since they tend to slightly increase with time due to the degradation of the spectrograph focus but still an increasing drift can be observed.}. The same happens to the H$\alpha$ periodogram, with a stronger peak at $\sim$\,392\,days and an alias at $\sim$\,357\,days also above the FAP level. This signal might be related to the period of 417 days observed in the photometry and be produced by rotational modulation of active regions. Finally, the BIS measurements show a peak just above the FAP line with exactly the same period as the \textit{RV} variability, 689.8\,days.

We also studied a possible relation between the \textit{RV} and line-profile indicators. In Fig. \ref{NGC2423_BIS} we can see how the \textit{RV} of NGC2423 No. 3 shows a strong correlation with biGauss and BIS (the Pearson's correlation coefficient is depicted in each plot). This fact warns us about the possibility that the signals we are observing are related to inhomogeneities in the stellar surface and not with an orbiting body. However, we note that the 18 \textit{RV} points initially presented in Paper I did not show any correlation with BIS and in some of the observing time windows the BIS is practically flat and thus not correlated with \textit{RV} (see Fig. \ref{ts_3_additional}). It is interesting however, that the sign of the slope for BIS vs \textit{RV} is opposite to what is found in many cases in the literature \citep[e.g.][]{queloz01,boisse09,figueira13}, as expected for \textit{RV} variations due to stellar spots. This case resembles that of the slow rotator HD\,41004A \citep{santos02} where BIS shows a strong positive correlation with \textit{RV}. In that system the best explanation for the \textit{RV} variation found and the positive correlation with BIS is given by the presence of a brown dwarf around the secondary component HD\,41004B (probably an M2 dwarf), which perturbs the contribution to the global CCF dominated by the primary. However, in our case, there are not reported binary companions around the planet candidate host but we cannot discard the possibility that there is a blended binary. By inspecting the SIMBAD database, we find that a foreground star from the cluster NGC2422 (PMS 419) with V=18.5 is at 9.18 arcsec distance from our star. That cluster is closer and younger than NGC2423 (490 pc vs 766 pc as in WEBDA database). By using the simulations done by \citet{cunha13} we can estimate a contamination of $\sim$1\,m\,s$^{-1}$ (considering that our target is a K5 star and the difference in apparent magnitude is 8). Thus, we can reject the hypothesis that such foreground star is affecting our \textit{RV} measurements. On the other hand we do not find a correlation of \textit{RV} with FWHM and contrast as also happens in the case of HD\,41004. Moreover, \citet{santerne15} simulated different blending scenarios (planet orbiting around a secondary unseen companion) and showed that when the secondary rotates faster than the primary (FWHM$_{1}$\,$<$\,FWHM$_{2}$) a positive correlation between BIS and \textit{RV} would be observed, as in our case. However, a positive correlation would also be observed for BiGauss and V$_{span}$ meanwhile here we do find a negative correlation for those two line profile indicators. Another case for a positive correlation between BIS and \textit{RV} was found for the active K giant EK Eri \citep{dall05} which later was reported to have a strong cool spot by \citet{auriere11} using the Zeeman-Doppler imaging method. 

The real value of \textit{v} sin \textit{i} is uncertain, as in the case of IC4651 No. 9122. Thus, it is difficult to relate the periodicity of the FWHM with rotational modulation of active regions but we cannot discard that possibility. Indeed, the different peaks observed in the periodograms of FWHM, H$\alpha$ and photometry (322, 392 and 417\,days), on one hand, and of BIS (689.8\,days) on the other hand, might indicate that the variations of the former are due to rotational modulation whereas the period of the latter is related to stellar pulsations which in turn can also be producing the periodic variation in \textit{RV}. Indeed, the fact of having a signal with eccentricity close to zero cannot rule out the possibility that pulsations are causing the \textit{RV} variation \citep{frink02}. For this star we find large peak-to-peak \textit{RV} amplitudes of $\sim$270\,m\,s$^{-1}$ meanwhile BIS variations are much smaller (40\,m\,s$^{-1}$), which could be a signature of oscillations \citep{hatzes96}.

In an attempt to further understand the different signal observed in this star we show the stacked BGLS periodogram of \textit{RV} in Fig. \ref{stacked_NGC2423}. The probability of the signal steadily increases by adding new observations, thus the signal is stable.

\section{A stellar signal disguising as a brown dwarf around NGC4349 No. 127} \label{sec:NGC4349}
In Paper I, a 19.8\,M$_{J}$ substellar companion around NGC4349 No. 127 was presented with a period of 677.8\,days, $e$\,=\,0.19, $K$\,=\,188\,m\,s$^{-1}$ and $a$\,=\,2.38\,UA. As for the previous case, this star is also the most evolved in the cluster \citep[see also Paper I and ][ for HR diagrams and \textit{RV} scatter histogram]{delgado16} and it clearly stands out with a \textit{RV} scatter of 70\,m\,s$^{-1}$ (the average \textit{RV} jitter is 20\,m\,s$^{-1}$). The estimated \textit{v} sin \textit{i} for this star is 6.1\,km\,s$^{-1}$ \citep[assuming a macrotubulence velocity of 2.1\,km\,s$^{-1}$ derived from][]{valenti05} and the radius is 37\,R$_\odot$ which leads to a maximum rotational period of $\sim$\,307\,days. However, as shown in previous sections, if we consider the macroturbulence given by \citet{hekker07} relation (4.8\,km\,s$^{-1}$) we would obtain \textit{v} sin \textit{i}\,=\,4.8\,km\,s$^{-1}$ and a maximum rotational period of $\sim$\,390\,days. The stellar parameters for this star can also be seen in Table \ref{tab:stellar}.

\subsection{Radial velocity dataset}

\begin{figure}%[h]
\centering
\includegraphics[width=1\linewidth]{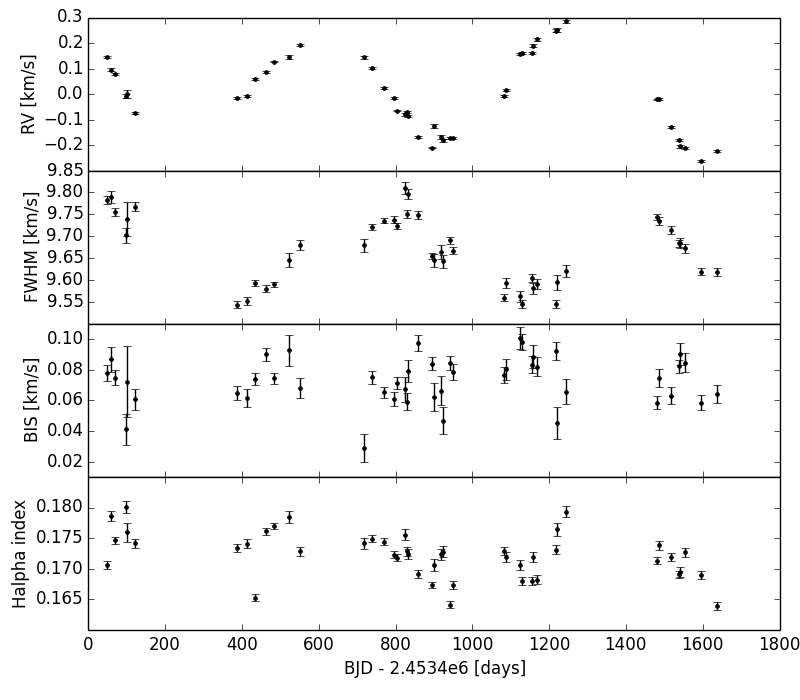}
\caption{Time-series of \textit{RV}, FWHM, BIS and H$\alpha$ index for NGC4349 No. 127.} 
\label{ts_127}
\end{figure}

\begin{figure}%[h]
\centering
\includegraphics[width=1\linewidth]{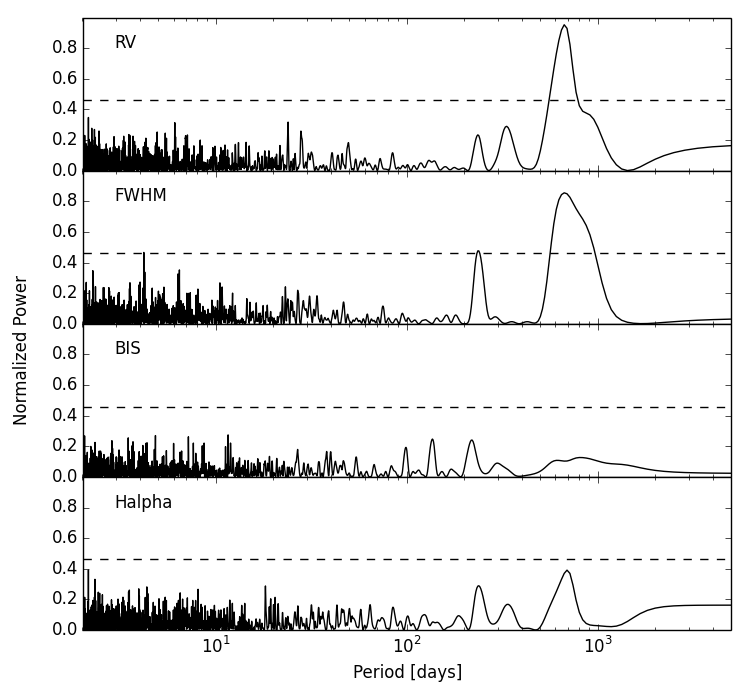}
\caption{Generalized Lomb-Scargle periodograms of \textit{RV}, FWHM, BIS and H$\alpha$ index for NGC4349 No. 127. The dashed line indicates the FAP at 1\% level.} 
\label{per_127}
\end{figure}

\begin{figure}
\centering
\includegraphics[width=1.0\linewidth]{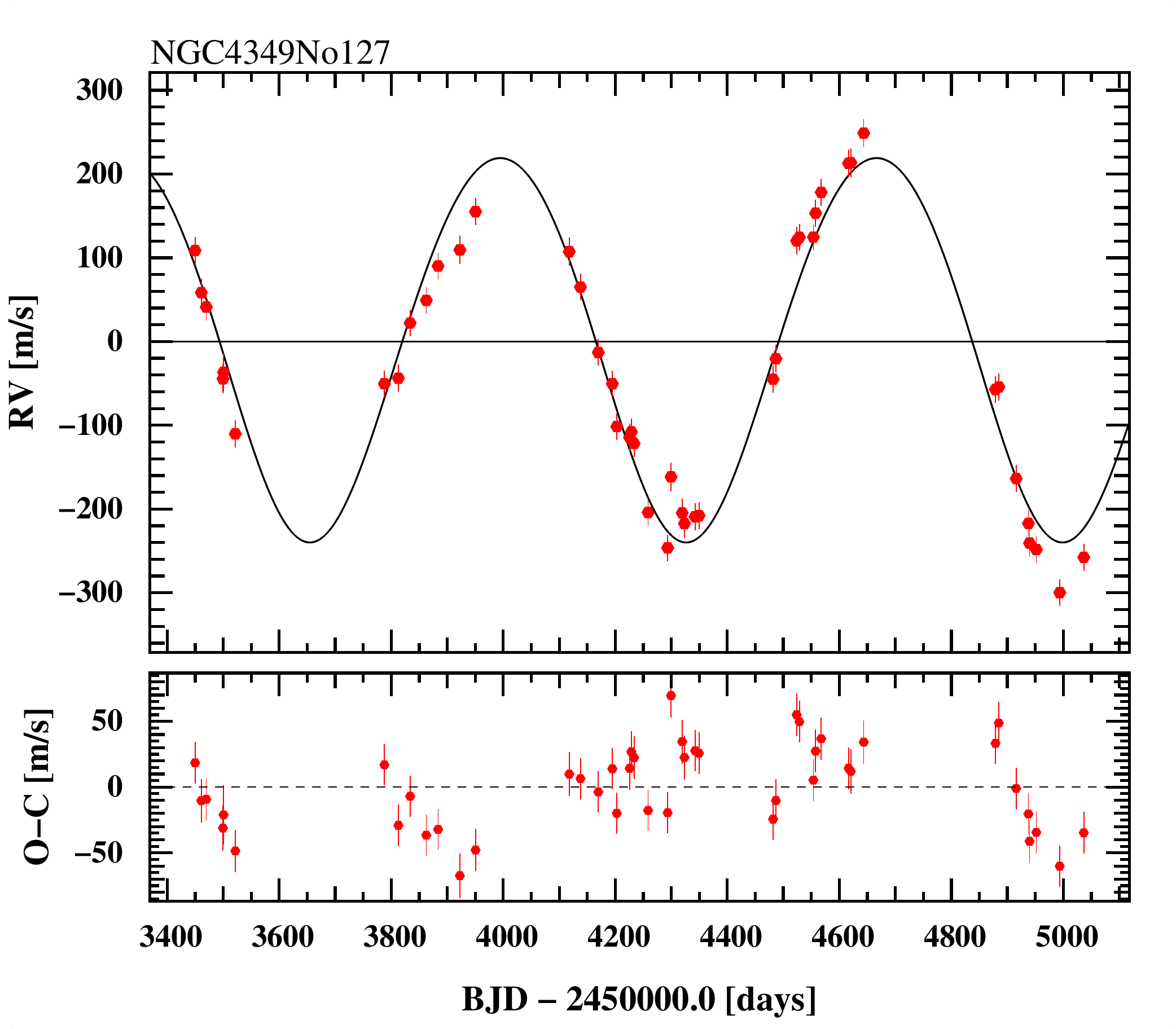}
\caption{Radial velocity data as a function of time for NGC4349 No. 127. The fitted orbit corresponds to a period of 672 days. A stellar jitter of 15\,m\,s$^{-1}$ has been added to the error bars.} 
\label{NGC4349_phase}
\end{figure}

In Fig. \ref{ts_127} we show the time series for \textit{RV}, FWHM, BIS and H$\alpha$ for NGC4349 No. 127. For this star the \ion{Na}{i} D$_1$ \& D$_2$ lines are contaminated. This plot includes 46 measurements collected during 1587 days while in Paper I, 20 \textit{RV} measurements along 784 days were presented. Figure \ref{per_127} shows the \textit{RV}, FWHM, BIS and H$\alpha$ periodograms for NGC4349 No. 127. The signal attributed to a brown dwarf candidate in Paper I has a significant peak at $\sim$666 days in the \textit{RV} periodogram which is similar to the period that we also find if we fit a Keplerian orbit to the data (672\,days, Fig. \ref{NGC4349_phase}). This value is very close to the 678 days period derived in Paper I. We obtain a planetary mass $m_{2}$ sin \textit{i}\,=\,24.1\,M$_{J}$ and the eccentricity value is 0.05, lower than previously reported value in Paper I. 

\subsection{Stellar activity, photometry and line profile analysis}

In Figure \ref{per_127} we can see how the FWHM periodogram also shows a very strong peak at $\sim$666 days (middle panel), indicating that there are variations in the star's atmosphere with the exact same period as the \textit{RV}. Moreover, the H$\alpha$ periodogram also shows a signal at a similar period, with $P$\,$\sim$\,689 days (however just below the FAP = 1\% level). We can also see a significant signal at 235\,days for the FWHM which has a counterpart in the H$\alpha$ periodogram but with a lower power.

The time-series of \textit{RV}, FWHM, and H$\alpha$ show clearly the signals detected in the respective periodograms (see Fig. \ref{per_127}). We note that the amplitude of the FWHM variation is significant (peak-to-peak variation of $\sim$\,250\,m s$^{-1}$). In the figure, we can observe that FWHM and \textit{RV} are not in phase. However, stellar rotationally modulated active regions are known to induce \textit{RV} signals not in phase with activity proxies \citep[e.g.][]{queloz01,santos14}. In their detection paper, \citet{lovis07} used BIS and the \ion{Ca}{ii} H\&K lines as activity diagnostic tools to validate the candidate brown dwarf. However, BIS loses sensitivity for slow rotators and the stellar flux in the \ion{Ca}{ii} H\&K spectral region for red stars is very low, delivering a low S/N for this activity index. This could explain why the authors missed the strong signals coming from the star's atmosphere with the same period as the observed \textit{RV} variations. 
  
\begin{figure}%[h]
\centering
\includegraphics[width=1\linewidth]{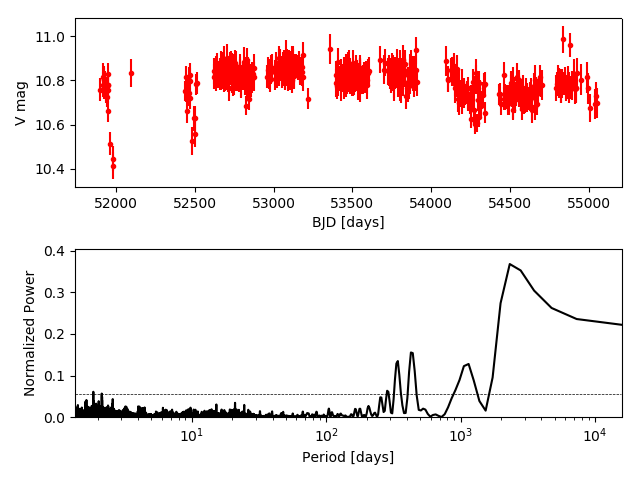}
\caption{GLS of V magnitude for NGC4349No127 using ASAS data. The dashed line indicates the FAP at 1\% level.} 
\label{NGC4349_phot}
\end{figure}

For this star we found two independent sets of photometric measurements from the ASAS-3 catalogue. In one of the sets we do not find any significant peak in the GLS applied to the light curve. The second set is shown in Fig. \ref{NGC4349_phot}. We find three significant signals in the periodogram with periods 428, 342 and 1149 days. The first two periods might be related to the rotational period of the star which we have estimated to be in between 300 and 400 days. However, there is no clear signal around the 666 days period found in the \textit{RV} although we note that the first estimation of the maximum rotational period of $\sim$\,307\,days and the signal in photometry at 342\,days are close to the first harmonic (\textit{P}/2) of the \textit{RV} period for NGC4349No127. 

On the other hand, we have compared the \textit{RV} with the line-profile indicators as done in previous section and we do not find significant correlations among the different indicators. Moreover, we show in Fig. \ref{stacked_NGC4349} the stacked BGLS for \textit{RV}. The signal looks stable and increase its significance by adding more observations which would be the expected behaviour for an orbiting body but such a trend is also expected if the stellar phenomena responsible for the \textit{RV} signal is stable over the time interval of our measurements.

\section{Discussion} \label{sec:discussion}

As we have seen in previous sections there can be different causes for the RV variability detected in our targets apart from the presence of planets. Long-period \textit{RV} variations with hundreds of days have been known to exist in several giant stars \citep[][]{walker89,larson93,hatzes93} with \textit{RV} amplitudes in the order of $\sim$50-400\,m\,s$^{-1}$ which were attributed to rotationally modulated active regions \citep{larson93,lambert87} or radial and non-radial pulsations \citep{hatzes99}, depending on the period of the signal. However, it is not clear yet whether pulsations (and of which kind) can create such high amplitude \textit{RV} variations as the ones observed. Indeed, stochastically-driven p-mode oscillations (also known as solar-type pulsations) are observed in giants from the lower RGB all the way to the AGB evolutionary phase \citep{mosser13}, where the most evolved of these are known as semi-regular variables. The properties of these pulsations follow specific scaling relations that can be used to predict their period and amplitude. Considering the stellar parameters estimated for the three stars under study, the expected periods would be less than a few days (the dynamical timescale for, e.g., NGC 2423 No. 3 is $\sim$\,20.8 hr). This, therefore, rules out the possibility that the variability seen in these stars is due to pulsations excited by a stochastic mechanism. On the other hand, if, by absurd, we were to assume that these type of pulsations justify the signal observed, and extrapolated from Eq. (9) of \citet{mosser13}, we would find an rms variability amplitude of 0.78 mag for modes with periods of about 698 days (the case of NGC2423 No. 3), hence a peak visible magnitude change of 1.1, which is well above what we see here. The use of the scaling relation in \citet{kjeldsen95} would then point to an rms velocity amplitude of the order of 10\,km\,s$^{-1}$, thus several orders of magnitude above what we see here.

Two other type of pulsations are observed in giants, that cannot be explained by a stochastically-driven mechanism. The first is the Mira variability, which shows much larger amplitudes than the oscillations discussed above. These are seen only at luminosities much higher than our target stars and, thus, can also be ruled out. The second - whose pulsation origin is still under debate \citep[e.g.][]{trabucchi17,takayama15} - has periods of several hundred days, significantly longer than the fundamental radial mode for the stars where they are observed. They appear as a particular sequence in the period-luminosity relation (sequence D) that is yet to be fully understood and are found both during the RGB and the AGB at luminosities  log\,(L/L$_{\odot}) \gtrsim$\,2.5\,dex. Recently they have been proposed to be manifestations of oscillatory convective modes \citep{saio15}. While in current theoretical models these appear only in giants with log\,(L/L$_{\odot}) \gtrsim$\,3\,dex (for a 2\,M$_{\odot}$ star), for models with reasonable convection parameters, the exact minimum luminosity depends on the mass and the mixing length adopted in the treatment of convection. NGC4349  No. 127 has log\,(L/L$_{\odot}) \sim$\,2.7 and could be a candidate for showing this type of pulsations, but we note that the period of 672 days is larger than expected for a typical sequence D star, unless the luminosity is significantly underestimated. The other two stars under study have much lower luminosity, so the presence of this type of pulsations seems to be even less likely, given the theoretical models. Also, their periods would be too long to fit into the sequence D described before. Moreover, the models by \citet{saio15} predict large surface temperature variations that are however not observed in our stars\footnote{We have derived \teff\ values for each individual spectrum and the maximum variation we observe is of $\sim$\,50\,K with no clear periodicity.}.

In summary, although pulsations are a possibility to explain the long-period variability seen in the RV, after considering the stellar parameters of the stars studied in this paper and the period and amplitude of the variability, we find that the type of pulsations discussed in the literature cannot comfortably explain our observations, with the possible exception of NGC4349 No. 127. We thus conclude, that either these are not produced by pulsations or they hint at a type of pulsation variability yet to be explained.

Alternatively, stellar activity and/or magnetic fields might be the cause for the \textit{RV} variability observed in giant stars. Among the three cases exposed in this work, probably, the easiest to explain would be NGC 4349 No. 127 since we find a perfect match between the periods of the \textit{RV}, the FWHM of the CCF and the H$_{\alpha}$ indicator. The variability we observe in these three indicators is thus presumably caused by rotational modulation of active regions. However, we are still far from fully understanding how the magnetic fields behave in giant stars and what are the timescales of stellar activity signals since their low rotation rates prevent their study through classical Doppler imaging, a successful technique to characterize the distribution of spots. Starspots with lifetimes of years have been observed with this technique in some stars \citep{strassmeier09,hussain02}. For example, polar cap-like starspots appear to have lifetimes of over a decade in RS CVn binaries as well as in young MS stars \citep{strassmeier09}. Nevertheless, a handful of giant stars with slow rotation have been studied in the last decade with the Zeeman Doppler imaging, which allows to obtain the strength and geometry of the surface magnetic field by using new high resolution spectropolarimeters such as Narval, ESPaDOnS or HARPSpol \citep{auriere15}. These authors found that most of the giant stars with detected magnetic fields (the longitudinal component) are undergoing the first dredge-up (i.e. they are close to the RGB base) and few of them are burning He in the core (red clump). Moreover, stars with rotational periods shorter than 200 days seem to have magnetic fields driven by a dynamo as in our Sun. Those stars present a good correlation between the strength of the magnetic field, the chromospheric activity ($S$-index) and the rotational period. On the other hand, stars such as EK Eri \citep{auriere11} or $\beta$ Ceti \citep{tsvetkova13} with $P_{rot}$ of 308.8 and 215 days, respectively, probably represent the prototype of descendants of magnetic Ap-Bp stars, hosting fossil magnetic fields. They have strong magnetic fields ($|B_{l}|_{max} >$\,10\,G) with lower than expected chromospheric activity. Finally, a last group of stars present very weak (sub Gauss level) but detectable magnetic fields with higher than expected chromospheric activity. Among this last group there are several interesting cases such as Pollux, Aldebaran and Arcturus.

Pollux ($\beta$ Gem) shows \textit{RV} variations with a period of 590 days spanning 25 years \citep{hatzes06} attributed to the presence of a planet. The star does not show any variability in the chromospheric \textit{S}-index or in the bisector velocity span \citep{hatzes06}. The photometric period does not match either that of the \textit{RV} variations. However, \textit{RV} values are correlated with the longitudinal magnetic field \citep{auriere09,auriere14}, casting doubts on the planetary nature of the signals. \citet{auriere15} also reported a weak magnetic field in Aldebaran ($\alpha$ Tau) but they could not determine the period of its variability. An extense study of \textit{RV} in this star spanning 20 years by \citet{hatzes15} reports the presence of a giant planet with a period of $\sim$\,630\,days and modulation of active regions with a period of $\sim$\,520\,days. Finally, a weak magnetic field was also detected in Arcturus ($\alpha$ Boo) by \citet{sennhauser11} with a possible period of 208\,days (only 3 measurements were made). The detection was confirmed by \citet{auriere15} but no period is provided. Therefore, it remains unclear if the high amplitude \textit{RV} variations (500\,m s$^{-1}$) with $P$\,=\,231 days detected by \citep{hatzes93} could be caused by the magnetic field. These three cases reflect very well the problematic of establishing the nature of different signals even when a large amount of data are collected during many years.

With the current data for the three stars studied here it is difficult to assure whether stellar chromospheric activity is affecting the \textit{RV} measurements. When adding all the spectra for a given star to increase the S/N in the blue region we cannot appreciate any sign of emission in the cores of \ion{Ca}{ii} H\&K lines. The semi-amplitude of the BIS has values of $\sim$20\,m s$^{-1}$ for IC4651 No. 9122 and NGC2423 No. 3 and $\sim$30\,m s$^{-1}$ for NGC4349 No. 127. Considering a maximum \textit{v} sin \textit{i} of 4\,km s$^{-1}$ for the former and 6\,km s$^{-1}$ for the latter, the presence of spots with a filling factor higher than 5\% would be needed to explain the BIS semi-amplitudes \citep[see Fig. 4 of ][]{santos03_planet}. However, for values of \textit{v} sin \textit{i} of 2\,km s$^{-1}$ or lower the presence of spots would not produce BIS variations. A spot filling factor of 5\%-10\% would produce a variability in $V$ of $\sim$0.05-0.11 mag which is below the peak-to-peak variability we observe for our stars (with approximate values of 0.3, 0.15 and 0.2\,mag for IC4651 No. 9122, NGC2423 No. 3 and NGC4349 No. 127, respectively). Therefore, the photometric variability we observe in our stars might be caused by spots (which in turn would produce the observed BIS variations) but the photometric periods do not match the period of the \textit{RV} neither the period of the BIS for any of the stars. 

We find striking the fact that the periods in \textit{RV} found for the three stars are all close to 700 days. We also find hints that the amplitude and the phase of the signal is changed along the time for the three cases. Curiosly, in a recent paper by \citet{hatzes18} they also report a \textit{RV} signal in $\gamma$ Draconis (a K giant with \teff\,=\,3990\,K, \logg\,=\,1.67, 2.14\,M$_\odot$ and L\,=\,510\,L$_{\odot}$) with a semi-amplitude of 148\,m s$^{-1}$ and a period of 702\,days that changes in phase and amplitude. It is interesting, though, that they do not find any correlation of \textit{RV} with the Ca II index or with the BIS. These authors suggest that the variability observed in $\gamma$ Draconis might be caused by dipole oscillatory convection modes \citep{saio15} which could be also the explanation for the \textit{RV} variability observed in NGC4349 No. 127 as discussed above. There are other cases in the literature with long-period \textit{RV} variations in M giants as well. For example, \citet{lee17} reported that the M giant HD\,36384 shows photometric variations with a period of 570\,days close to the period of 535\,days in the \textit{RV}. Furthermore, the M giant $\mu$ UMa shows \textit{RV} variations with a period of 471\,days with bisector velocity curvature variations of 463\,days \citep{lee16}. Moreover, the EWs of H$\alpha$ and H$\beta$ lines show periodic variations with 473\,days for this star. Recently, \citet{bang18} found \textit{RV} variations with a period of 719\,days in the M giant HD\,18438 which also shows photometric and H$\alpha$ variability with similar period. All these examples in the literature together with the three stars presented here warns us about the difficulty of confirming the planetary nature of periodic \textit{RV} variations in red giants, specially the most massive ones.

In the last years several works have discussed the planet occurrence as a function of stellar mass finding different results \citep[e.g.][]{omiya09,johnson10,reffert15,jones17}. When Paper I was published, only 6 substellar companions\footnote{Nowadays, this number goes up to 67 companions by using the data collected in The Extrasolar Planets Encyclopaedia} were known around stars more massive than $\sim$\,1.78\,M$_\odot$ and they presented much higher masses on average than for lower mass stars (see Fig. 11 in Paper I), leading to the hypothesis that the frequency of massive planets is higher around more massive stars. Certainly, the inclusion of the brown dwarf around the massive star NGC4349 No. 127 contributed to such correlation. Later studies by the Korean--Japanese planet search program confirmed such result \citep{omiya12}. The works by \citet{reffert15,jones16} also suggest that the planet occurrence rate increases with stellar mass but peaks around 2\,M$_\odot$ and sharply decreases for stars more massive than 2.5-3\,M$_\odot$ in agreement with the theoretical model by \citet{kennedy08}. Indeed, the 12-years length Lick Observatory survey \citep{reffert15} did not find any planet around stars more massive than 2.7\,M$_\odot$ despite having a significant number of stars with masses up to 5\,M$_\odot$. Similarly, the 6-years length EXPRESS survey has not found any planet in stars more massive than 2.5\,M$_\odot$ \citep{jones16}. Nowadays, in the Extrasolar Planets Encyclopaedia we can find only 8 stars with masses above 2.7\,M$_\odot$ hosting planets or brown dwarfs\footnote{We note that not all the planet hosts have an entry for the stellar mass in this database and the available values listed there come from different sources/methods, thus any conclusion made from this heterogeneous dataset must be taken with caution.}, being HD\,13189 the only one more massive than NGC4349 No. 127, which should now be excluded. Moreover, the two bodies system around the 2.8\,M$_\odot$ star BD+20 2457 seems to be dynamically unstable \citep{horner14}, thus, the determination of planet occurrence rates with surveys covering the high stellar mass regime is advisable. As mentioned in the introduction, the measurement of masses for giant stars is a difficult task and several works discuss about the suitability of using isochrones compared with other methods such as asteroseismology which delivers much more precise masses. In order to do a proper statistical analysis evaluating planet occurrence rates we need to have a sample of homogeneously derived masses. Such study is out of the scope of the present paper but will be approached in a future work when the data collection for our survey in open clusters is completed.

\section{Conclusions} \label{sec:conclusion}

In this work we present the \textit{RV} data for three stars within the $\sim$\,15\,years survey of radial velocities in a sample of more than 142 giant stars in 17 open clusters. Additional data was collected during the last year. The large span of the data is needed in order to find long period planets since the probability of discovering short period planets around such kind of stars diminishes as the stars evolve and expand their radii. The long-term monitoring of giant stars is also essential to distinguish \textit{RV} variations caused by the rotational modulation of active regions because the rotational periods of evolved stars are much longer than for dwarfs. 

We report the possible discovery of a 7.2\,M$_{J}$ planet in a 747 days orbit around the star No. 9122 in the intermediate-age open cluster IC4651. If confirmed, the planet would be at a distance of 2.05 AU and the eccentricity of the orbit would be 0.15. However, we find that the FWHM values of the CCF and the H$\alpha$ index show a periodic behavior with a bit shorter period than the \textit{RV} values. The period of the FWHM and H$\alpha$ index variation get even closer to that of the \textit{RV} when adding few more points taken after 2009. Nevertheless, we note that the significance of the periodic signals on those two indicators is low. We perform further tests on this system to understand whether the periodic \textit{RV} signals are of stellar origin. Although there exists the possibility of having both rotational modulation and a planet with close periods we conclude that more data will be needed to validate or refute the presence of such planet.

This survey already reported two substellar companions in Paper I, NGC2423 No. 3b and NGC4349 No. 127b. These stars were further followed in the next two years after the publication of Paper I (for NGC2423 No. 3 additional observations were carried out during last semester) and here we present an updated analysis of the \textit{RV} data. We find that the \textit{RV} variability of NGC2423 No. 3 is strongly correlated with the BIS of the CCF when using the full dataset, a fact that by itself probably invalidates the planet presence although we cannot explain the reason for such correlation. However, we note that such correlation is not stable along the time and the BIS becomes flat in some of the observed time windows. On the other hand, for NGC4349 No. 127 we find that the FWHM of the CCF presents a large amplitude periodic variation with the same strong peak as the \textit{RV}. Moreover, the periodogram of the H$\alpha$ index also shows a peak with a similar period very close to the 1\% FAP level. Therefore, we believe that the presence of active regions in the atmosphere and not a substellar companion is responsible for the \textit{RV} variability. For this star, the presence of a new kind of stellar pulsations might be a plausible explanation for the observed variability as well.

Long-term period signals have been observed in many K giants either produced by stellar activity or by oscillations. The amplitudes of these signals can also reach hundreds of m\,s$^{-1}$, mimicking the presence of giant planets. The three cases presented in this work are a clear example of the need to perform detailed analysis of \textit{RV} modulations in giant stars before the real nature of a planetary signal can be assumed. Moreover, the analysis of the FWHM of the CCF has revealed as essential in order to discard the presence of a companion but we note that this indicator is not widely used in the validation of planets around evolved stars. Finally, more theoretical and observational study is needed to further understand the different nature of stellar oscillations in evolved stars.

\begin{acknowledgements}
We thank Fran\c{c}ois Bouchy and Xavier Dumusque for coordinating the shared observations with HARPS and all the observers who helped collecting the data.  

E.D.M., V.A., P.F., N.C.S., S.G.S. and J. P. F. acknowledge the support from Funda\c{c}\~ao para a Ci\^encia e a Tecnologia (FCT) through national funds and from FEDER through COMPETE2020 by the following grants: UID/FIS/04434/2013 \& POCI--01--0145-FEDER--007672, PTDC/FIS-AST/1526/2014 \& POCI--01--0145-FEDER--016886, PTDC/FIS-AST/7073/2014 \& POCI-01-0145-FEDER-016880 and POCI-01-0145-FEDER-028953.
E.D.M. acknowledges the support from FCT through Investigador FCT contract IF/00849/2015/CP1273/CT0003 and in the form of an exploratory project with the same reference. V.A., P.F., N.C.S., S.G.S. and M.C. also acknowledge the support from FCT through Investigador FCT contracts IF/00650/2015/CP1273/CT0001, IF/01037/2013/CP1191/CT0001, IF/00169/2012/CP0150/CT0002, IF/00028/2014/CP1215/CT0002, and IF/00894/2012/CP0150/CT0004. J.P.F. acknowledges support by the fellowships SFRH/BD/93848/2013 funded by FCT (Portugal) and POPH/FSE (EC). PF further acknowledges support from Funda\c{c}\~ao para a Ci\^encia e a Tecnologia (FCT) in the form of an exploratory project of reference IF/01037/2013/CP1191/CT0001. AM acknowledges funding from the European Union Seventh Framework Programme (FP7/2007-2013) under grant agreement number 313014 (ETAEARTH).

This research has made use of the The Extrasolar Planets Encyclopaedia, SIMBAD and WEBDA databases. This work has also made use of the IRAF facility. 

\end{acknowledgements}
\vspace{-0.4cm}

%----------------------------------------------------------------------------------------
%       Bibliography
%----------------------------------------------------------------------------------------
\bibliography{edm_bibliography_planets}

%----------------------------------------------------------------------------------
%       Appendices
%----------------------------------------------------------------------------------

\appendix

\section{Additional figures}
%----------------------------------------- FIGURE ---------------------------------------------------
\begin{figure}%[h]
\centering
\includegraphics[width=6cm]{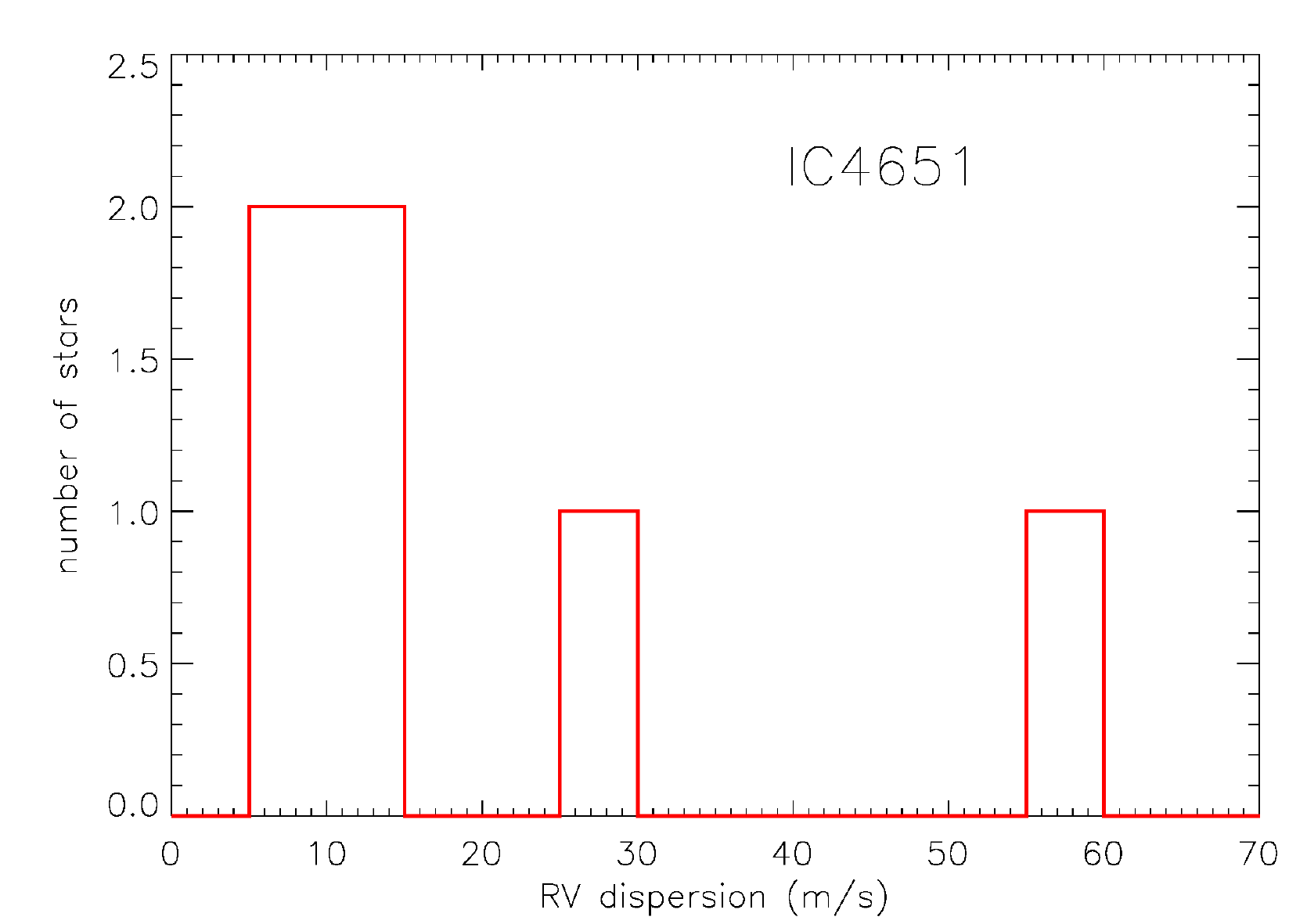}
\caption{Standard deviation of \textit{RV} for members in the cluster IC4651. Two stars with dispersions higher than 100\,m\,s$^{-1}$ (due to stellar or brown dwarf companions) are not shown here.} 
\label{IC4651_scatter}
\end{figure}
%--------------------------------------------------------------------------------------------------

\begin{figure}[h]
\centering
\includegraphics[width=1\linewidth]{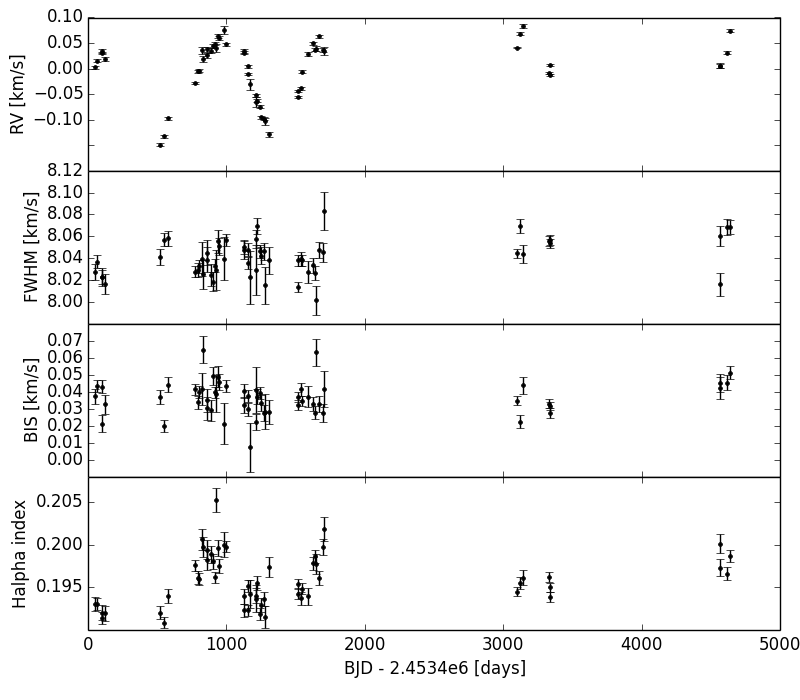}
\caption{Time series of \textit{RV}, FWHM, BIS and H$\alpha$ index for IC4651 No. 9122.} 
\label{ts_9122_additional}
\end{figure}

\begin{figure}%[h]
\centering
\includegraphics[width=1\linewidth]{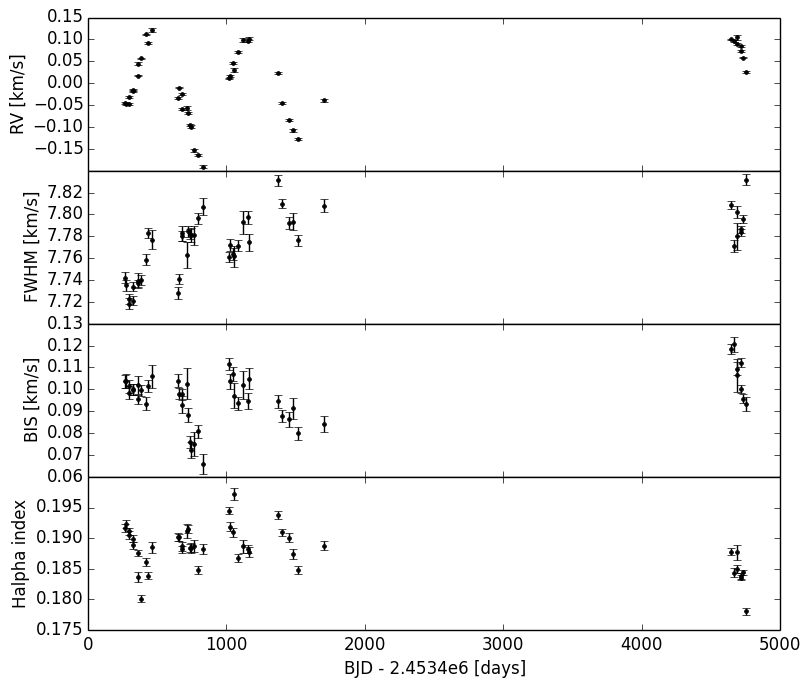}
\caption{Time series of \textit{RV}, FWHM, BIS and H$\alpha$ index for NGC2423 No. 3.} 
\label{ts_3_additional}
\end{figure}

\begin{figure}
\centering
\includegraphics[width=0.7\linewidth]{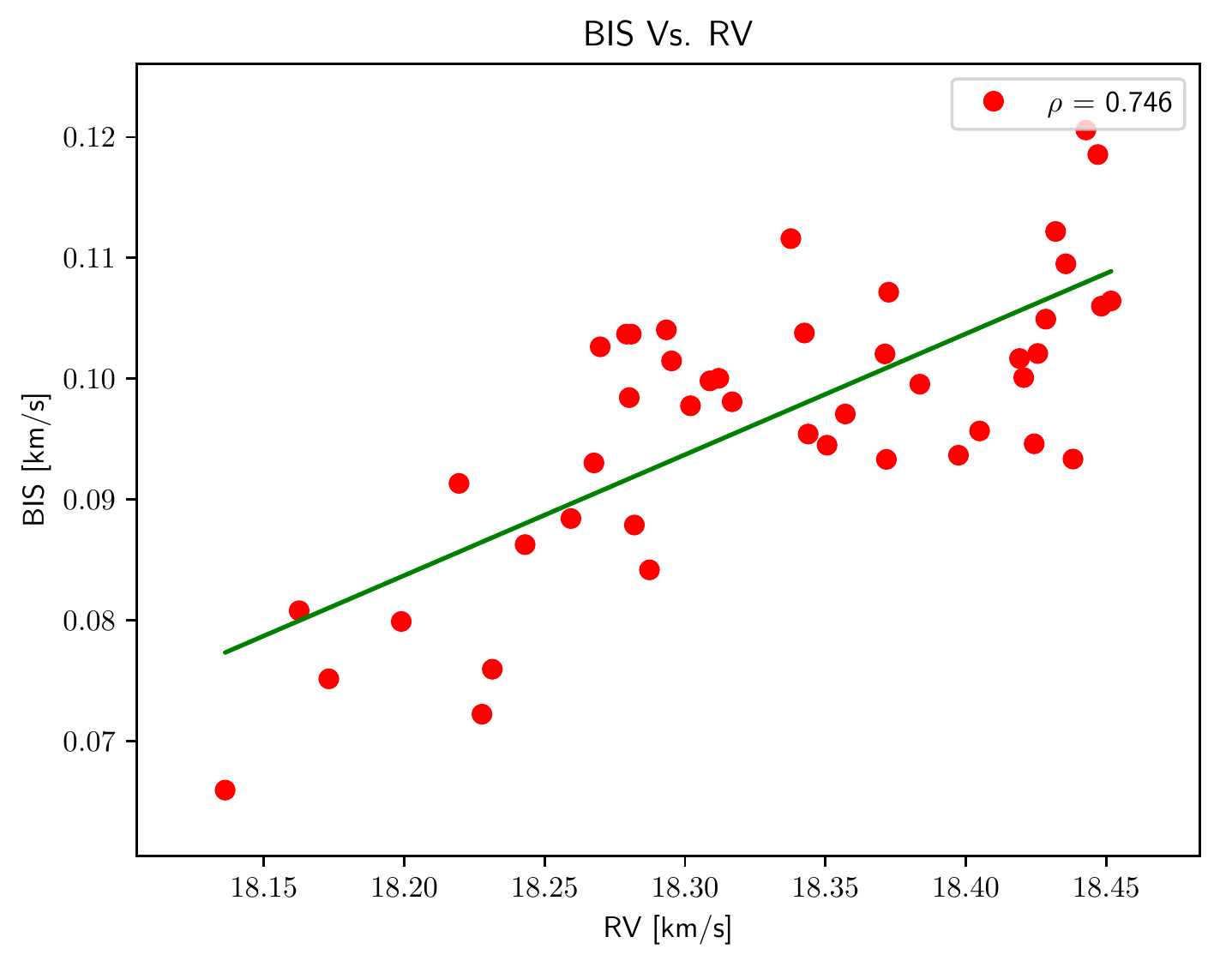}
\includegraphics[width=0.7\linewidth]{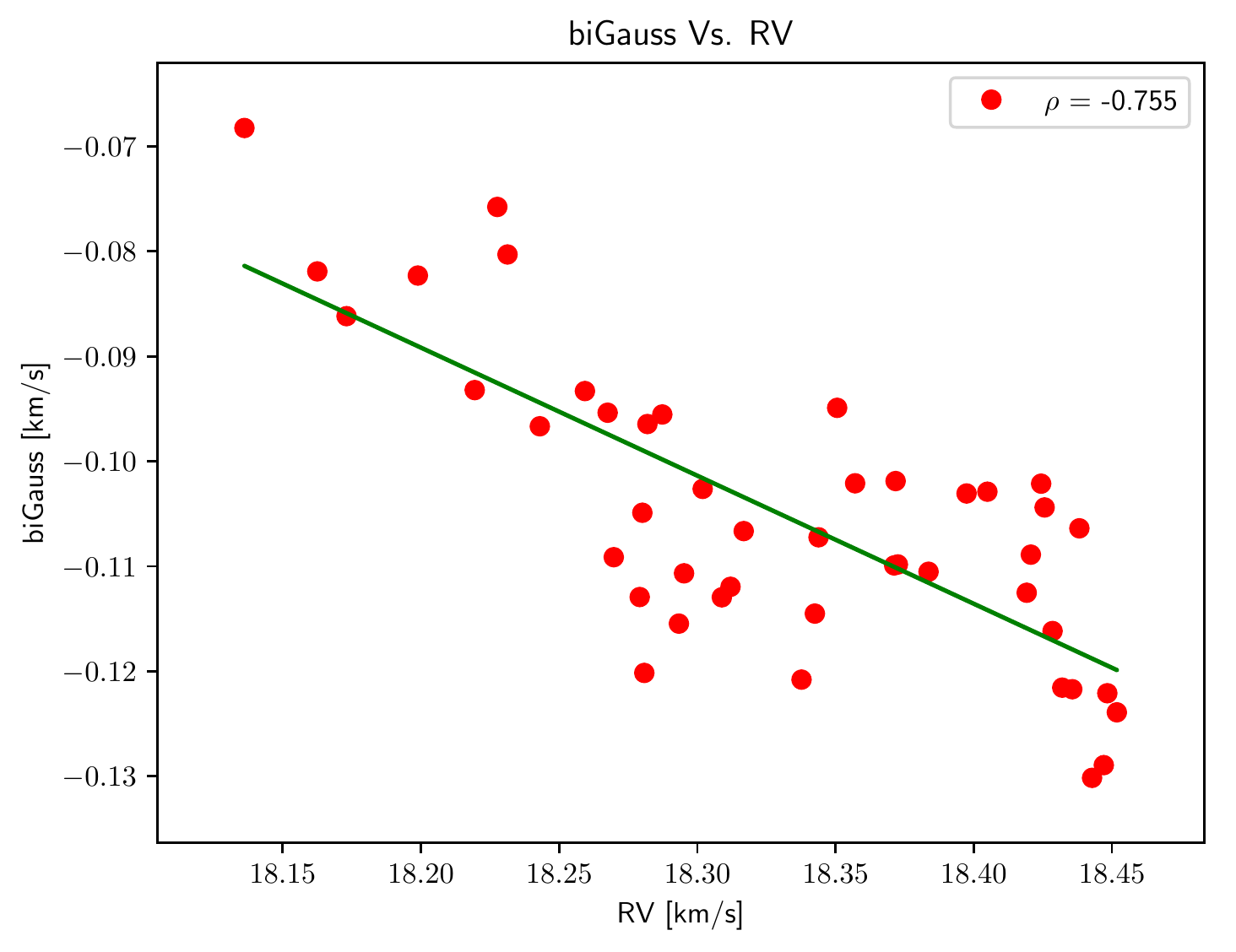}
\caption{Correlation between \textit{RV} and the line profile indicators BIS and biGauss for NGC2423 No. 3.}
\label{NGC2423_BIS}
\end{figure}

\begin{figure}%[h]
\centering
\includegraphics[width=1\linewidth]{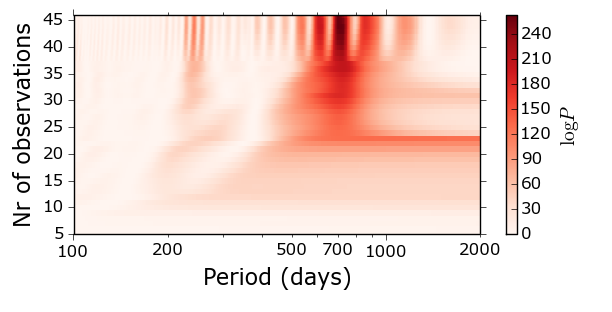}
\caption{Stacked periodogram for \textit{RV} measurements in NGC2423 No. 3.} 
\label{stacked_NGC2423}
\end{figure}

\begin{figure}%[h]
\centering
\includegraphics[width=1\linewidth]{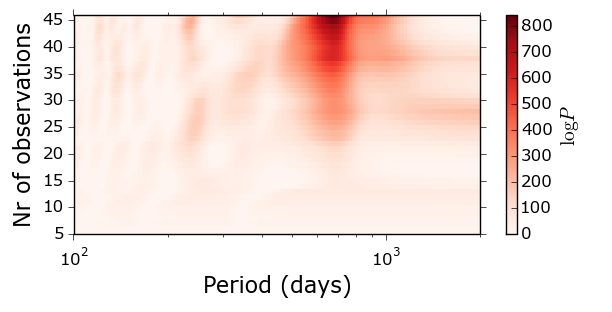}
\caption{Stacked periodogram for \textit{RV} measurements in NGC 4349 No. 127.} 
\label{stacked_NGC4349}
\end{figure}

\section{\textit{RV} measurements}\label{app:measurements}

\begin{center}
\begin{table}[h!]
% use packages: array
\caption{Radial velocity measurements for IC4651 No. 9122 obtained
with HARPS and their instrumental error bars. All data are relative to
the solar system barycenter.}
\centering
\begin{tabular}{lcc}
\hline
\noalign{\smallskip} 
JD-2\,400\,000 &  \textit{RV} & Uncertainty \\
\noalign{\smallskip} 
     & [km\,s$^-1$] & [km\,s$^-1$]  \\
\noalign{\smallskip} 
\hline
\hline
\noalign{\smallskip} 
53449.887674  &  -30.25225  & 0.00317   \\
53462.850455  &  -30.23935  & 0.00257   \\
53499.742442  &  -30.22078  & 0.00363   \\
53500.698475  &  -30.22416  & 0.00272   \\
53521.693417  &  -30.23617  & 0.00397   \\
53921.668152  &  -30.40372  & 0.00302   \\
53950.621071  &  -30.38763  & 0.00256   \\
53981.622808  &  -30.35229  & 0.00301   \\
54171.886988  &  -30.28285  & 0.00231   \\
54194.857714  &  -30.25975  & 0.00282   \\
54200.849131  &  -30.25867  & 0.00267   \\
54226.894524  &  -30.21881  & 0.00658   \\
54232.731181  &  -30.23582  & 0.00561   \\
54256.735797  &  -30.21726  & 0.00492   \\
54261.807071  &  -30.22907  & 0.00488   \\
54291.722147  &  -30.21941  & 0.00428   \\
54299.690631  &  -30.20983  & 0.00379   \\
54319.614198  &  -30.20572  & 0.00244   \\
54323.579256  &  -30.21409  & 0.00786   \\
54342.555162  &  -30.19083  & 0.00437   \\
54349.595115  &  -30.19470  & 0.00322   \\
54385.490351  &  -30.17916  & 0.00851   \\
54393.518352  &  -30.20727  & 0.00239   \\
54523.892742  &  -30.22074  & 0.00341   \\
54529.878124  &  -30.22392  & 0.00284   \\
54553.859622  &  -30.25017  & 0.00255   \\
54556.860227  &  -30.26548  & 0.00240   \\
54568.915197  &  -30.28588  & 0.01020   \\
54615.900828  &  -30.32074  & 0.00974   \\
54616.851591  &  -30.30711  & 0.00347   \\
54621.885865  &  -30.31795  & 0.00317   \\
54641.777674  &  -30.32915  & 0.00254   \\
54646.614534  &  -30.35011  & 0.00332   \\
54672.612166  &  -30.35315  & 0.00332   \\
54679.582438  &  -30.35829  & 0.00721   \\
54708.574369  &  -30.38375  & 0.00517   \\
54913.858276  &  -30.31120  & 0.00213   \\
54919.803228  &  -30.29862  & 0.00200   \\
54936.812674  &  -30.29323  & 0.00286   \\
54949.768030  &  -30.26058  & 0.00211   \\
54990.708393  &  -30.22648  & 0.00427   \\
55024.810993  &  -30.20521  & 0.00291   \\
55038.738466  &  -30.21751  & 0.00282   \\
55045.784236  &  -30.21564  & 0.00559   \\
55072.564309  &  -30.19168  & 0.00319   \\
55095.574817  &  -30.21826  & 0.00370   \\
55105.500100  &  -30.21990  & 0.00749   \\                    
56499.579896  &  -30.21348  & 0.00184   \\
56521.628162  &  -30.18630  & 0.00278   \\
56543.505489  &  -30.17094  & 0.00352   \\
56733.852100  &  -30.26437  & 0.00198   \\
56734.867908  &  -30.24730  & 0.00200   \\
56735.863877  &  -30.26788  & 0.00173   \\
57968.650782  &  -30.25014  & 0.00400   \\
57969.606900  &  -30.24899  & 0.00450   \\
58013.569803  &  -30.22394  & 0.00317   \\
58041.498430  &  -30.18038  & 0.00278   \\   
\hline
\noalign{\medskip} %
\end{tabular}
\end{table}
\label{IC4651_rv}
\end{center}

\begin{center}  
\begin{table}[h!]
\caption{Radial velocity measurements for NGC2423 No. 3 obtained
with HARPS and their instrumental error bars taken afte JD-2\,454\,130. Older data is presented in Paper I. All data are relative to
the solar system barycenter.}
\centering      
\begin{tabular}{lcc}
\hline          
\noalign{\smallskip} 
JD-2\,400\,000 &  \textit{RV} & Uncertainty \\
\noalign{\smallskip} 
     & [km\,s$^-1$] & [km\,s$^-1$]  \\
\noalign{\smallskip} 
\hline
\hline
\noalign{\smallskip} 
53669.846074   &  18.28083  &  0.00215   \\
53674.796790   &  18.27917  &  0.00224   \\
53692.862440   &  18.28013  &  0.00217   \\
53699.842505   &  18.29520  &  0.00210   \\
53721.855913   &  18.30887  &  0.00187   \\
53728.752333   &  18.31202  &  0.00191   \\
53758.656817   &  18.37122  &  0.00301   \\
53764.700951   &  18.34386  &  0.00150   \\
53784.639237   &  18.38365  &  0.00190   \\
53817.563704   &  18.43820  &  0.00214   \\
53831.568254   &  18.41915  &  0.00195   \\
53861.565085   &  18.44831  &  0.00369   \\
54050.824982   &  18.29334  &  0.00226   \\
54054.864236   &  18.31679  &  0.00191   \\
54078.820691   &  18.30194  &  0.00190   \\
54082.782001   &  18.26757  &  0.00288   \\
54114.702865   &  18.26979  &  0.00499   \\
54122.715976   &  18.25936  &  0.00235   \\
54138.717872   &  18.23135  &  0.00197   \\
54142.652470   &  18.22766  &  0.00270   \\
54166.590019   &  18.17312  &  0.00385   \\
54197.574062   &  18.16256  &  0.00218   \\
54228.532429   &  18.13619  &  0.00333   \\
54421.769842   &  18.33770  &  0.00209   \\
54428.783945   &  18.34252  &  0.00248   \\
54447.806766   &  18.37255  &  0.00234   \\
54454.813715   &  18.35709  &  0.00399   \\
54482.750391   &  18.39740  &  0.00220   \\
54522.713892   &  18.42563  &  0.00452   \\
54554.593644   &  18.42438  &  0.00256   \\
54563.523609   &  18.42850  &  0.00332   \\
54775.842423   &  18.35059  &  0.00208   \\
54801.736645   &  18.28197  &  0.00193   \\
54853.716032   &  18.24304  &  0.00251   \\
54884.669017   &  18.21949  &  0.00333   \\
54916.569668   &  18.19892  &  0.00213   \\
55106.856747   &  18.28734  &  0.00267   \\
58043.863734   &  18.42718  &  0.00175   \\
58068.809819   &  18.42294  &  0.00247   \\
58089.731546   &  18.43190  &  0.00539   \\
58089.843208   &  18.41577  &  0.00232   \\
58116.705019   &  18.40079  &  0.00152   \\
58117.703625   &  18.41215  &  0.00157   \\
58129.814248   &  18.38507  &  0.00160   \\
58156.798523   &  18.35187  &  0.00224   \\                           
\hline
\noalign{\medskip} %
\end{tabular}
\end{table}
\label{NGC2423_rv_harps}
\end{center}

\begin{center}
\begin{table}[h!]
\caption{Radial velocity measurements for NGC4349 No. 127 obtained
with HARPS and their instrumental error bars taken after JD-2\,454\,234. Older data is presented in Paper I. All data are relative to
the solar system barycenter.}
\centering
\begin{tabular}{lcc}
\hline
\noalign{\smallskip} 
JD-2\,400\,000 &  \textit{RV} & Uncertainty \\
\noalign{\smallskip} 
     & [km\,s$^-1$] & [km\,s$^-1$]  \\
\noalign{\smallskip} 
\hline
\hline
\noalign{\smallskip} 
54258.557903  &  -11.63628  &   0.00377  \\
54293.529231  &  -11.67849  &   0.00298  \\
54299.560888  &  -11.59373  &   0.00653  \\
54319.473660  &  -11.63690  &   0.00664  \\
54323.471152  &  -11.64945  &   0.00630  \\
54342.475167  &  -11.64134  &   0.00325  \\
54349.471386  &  -11.64002  &   0.00366  \\
54481.827906  &  -11.47722  &   0.00356  \\
54486.798601  &  -11.45261  &   0.00484  \\
54523.803004  &  -11.31187  &   0.00511  \\
54528.781327  &  -11.30763  &   0.00382  \\
54553.727450  &  -11.30752  &   0.00384  \\
54557.719105  &  -11.27895  &   0.00531  \\
54567.681649  &  -11.25400  &   0.00460  \\
54616.604433  &  -11.21941  &   0.00421  \\
54620.561443  &  -11.21867  &   0.00733  \\
54643.591230  &  -11.18338  &   0.00576  \\
54878.815480  &  -11.48937  &   0.00312  \\
54884.800643  &  -11.48642  &   0.00419  \\
54915.765632  &  -11.59573  &   0.00384  \\
54937.688513  &  -11.64931  &   0.00312  \\
54939.702640  &  -11.67278  &   0.00507  \\
54951.725335  &  -11.68053  &   0.00442  \\
54993.573342  &  -11.73192  &   0.00349  \\
55036.503475  &  -11.68980  &   0.00401  \\
\hline
\noalign{\medskip} 
\end{tabular}
\end{table}
\label{NGC4349_rv}
\end{center}

\clearpage

\end{document}